\documentclass[prd,superscriptaddress,twocolumn,amssymb,amsmath,amsfonts,aps,nofootinbib]{revtex4-1}

\newcommand{\gx}{\textsc{GlueX}}

\usepackage{graphicx}
\usepackage[colorlinks=true,allcolors=blue,bookmarks=true]{hyperref}

\begin{document}

\title{\boldmath A study of meson and baryon decays to strange final states with \gx~in Hall D \\ 
{\small  (A proposal to the 39$^\mathrm{th}$ Jefferson Lab Program Advisory Committee)}}

\date{May 3, 2012}

\author{M.~Dugger}
\author{B.~Ritchie}
\affiliation{Arizona State University, Tempe, Arizona 85287, USA}
\author{E.~Anassontzis}
\author{P.~Ioannou}
\author{C.~Kourkoumeli}
\author{G.~Voulgaris}
\affiliation{University of Athens, GR-10680 Athens, Greece}
\author{N.~Jarvis}
\author{W.~Levine}
\author{P.~Mattione}
\author{C.~A.~Meyer}\thanks{Spokesperson}
\author{R.~Schumacher}
\affiliation{Carnegie Mellon University, Pittsburgh, Pennsylvania 15213, USA}
\author{P.~Collins}
\author{F.~Klein}
\author{D.~Sober}
\affiliation{Catholic University of America, Washington, D.C.\ 20064, USA}
\author{D.~Doughty}
\affiliation{Christopher Newport University, Newport News, Virginia 23606, USA}
\author{A.~Barnes}
\author{R.~Jones}
\author{J.~McIntyre}
\author{F.~Mokaya}
\author{B.~Pratt}
\author{I.~Senderovich}
\affiliation{University of Connecticut, Storrs, Connecticut 06269, USA}
\author{W.~Boeglin}
\author{L.~Guo}
\author{P.~Khetarpal}
\author{E.~Pooser}
\author{J.~Reinhold}
\affiliation{Florida International University, Miami, Florida 33199, USA}
\author{H.~Al Ghoul}
\author{S.~Capstick}
\author{V.~Crede}
\author{P.~Eugenio}
\author{A.~Ostrovidov}
\author{N.~Sparks}
\author{A.~Tsaris}
\affiliation{Florida State University, Tallahassee, Florida 32306, USA}
\author{D.~Ireland}
\author{K.~Livingston}
\affiliation{University of Glasgow, Glasgow G12 8QQ, United Kingdom}
\author{D.~Bennett}
\author{J.~Bennett}
\author{J.~Frye}
\author{J.~Leckey}
\author{R.~Mitchell}
\author{K.~Moriya}
\author{M.~R.~Shepherd}\thanks{Deputy Spokesperson}
\author{A.~Szczepaniak}
\affiliation{Indiana University, Bloomington, Indiana 47405, USA}
\author{R.~Miskimen}
\affiliation{University of Massachusetts, Amherst, Massachusetts 01003, USA}
\author{M.~Williams}
\affiliation{Massachusetts Institute of Technology, Cambridge, Massachusetts 02139, USA}
\author{P.~Ambrozewicz}
\author{A.~Gasparian}
\author{R.~Pedroni}
\affiliation{North Carolina A\&T State University, Greensboro, North Carolina 27411, USA}
\author{T.~Black}
\author{L.~Gan}
\affiliation{University of North Carolina, Wilmington, North Carolina 28403, USA}
\author{J.~Dudek}
\affiliation{Old Dominion University, Norfolk, Virginia 23529, USA}
\affiliation{Thomas Jefferson National Accelerator Facility, Newport News, Virginia 23606, USA}
\author{F.~Close}
\affiliation{University of Oxford, Oxford OX1 3NP, United Kingdom}
\author{E.~Swanson}
\affiliation{University of Pittsburgh, Pittsburgh, Pennsylvania 15260, USA }
\author{S.~Denisov}
\affiliation{Institute for High Energy Physics, Protvino, Russia}
\author{G.~Huber}
\author{S.~Katsaganis}
\author{D.~Kolybaba}
\author{G.~Lolos}
\author{Z.~Papandreou}
\author{A.~Semenov}
\author{I.~Semenova}
\author{M.~Tahani}
\affiliation{University of Regina, Regina, SK S4S 0A2, Canada}
\author{W.~Brooks}
\author{S.~Kuleshov}
\author{A.~Toro}
\affiliation{Universidad T\'ecnica Federico Santa Mar\'ia, Casilla 110-V Valpara\'iso, Chile}
\author{F.~Barbosa}
\author{E.~Chudakov}\thanks{Hall D Leader}
\author{H.~Egiyan}
\author{M.~Ito}
\author{D.~Lawrence}
\author{L.~Pentchev}
\author{Y.~Qiang}
\author{E.~S.~Smith}
\author{A.~Somov}
\author{S.~Taylor}
\author{T.~Whitlatch}
\author{E.~Wolin}
\author{B.~Zihlmann}
\affiliation{Thomas Jefferson National Accelerator Facility, Newport News, Virginia 23606, USA}

\collaboration{The \gx~Collaboration}

\begin{abstract}

The primary motivation of the \gx~experiment is to search for and ultimately study the pattern of
gluonic excitations in the meson spectrum produced in $\gamma p$ collisions.  
Recent lattice QCD calculations predict a rich spectrum
of hybrid mesons that have both exotic and non-exotic $J^{PC}$, corresponding to $q\bar{q}$ ($q=u,$ $d,$ or $s$)
states coupled with a gluonic field.  A thorough study of the hybrid spectrum, including the identification
of the isovector triplet, with charges 0 and $\pm1$, and both isoscalar members, 
$|s\bar{s}\rangle$ and $|u\bar{u}\rangle + |d\bar{d}\rangle$, for each predicted hybrid combination of $J^{PC}$, may only be 
achieved by conducting a systematic amplitude analysis of many different hadronic final states.  We propose
the development of a kaon identification system, supplementing the existing \gx~forward time-of-flight
detector, in order to cleanly select meson and baryon decay channels that include kaons.  Once this detector
has been installed and commissioned, we plan to collect a total of 200 days of physics analysis data at an 
average intensity of $5\times 10^7$ tagged photons on target per second.  This data sample will provide an order of 
magnitude statistical improvement over the initial \gx~data set and, with the developed kaon identification system, 
a significant increase in the potential for \gx~to make key experimental advances in our knowledge of 
hybrid mesons and $\Xi$ baryons.

\end{abstract}

\maketitle

\section{Introduction and background}

A long-standing goal of hadron physics has been to understand how the quark 
and gluonic degrees of freedom that are present in the fundamental QCD Lagrangian 
manifest themselves in the spectrum of hadrons.   Of particular interest is how the 
gluon-gluon interactions might give rise to physical states with gluonic excitations.  
One class of such states is the hybrid mesons, 
which can be naively thought of as quark anti-quark pairs coupled to a valence 
gluon ($q\bar{q}g$).  Recent lattice QCD calculations~\cite{Dudek:2011bn} predict a rich spectrum 
of hybrid mesons.  A subset of these hybrids have an unmistakable experimental signature:  
angular momentum ($J$), parity ($P$), and charge conjugation ($C$) that cannot be 
created from just a quark-antiquark pair.  Such states are called exotic hybrid mesons.  
The primary goal of the \gx~experiment in Hall~D is to search for and study these mesons.

A detailed overview of the motivation for the \gx~experiment as well as the design 
of the detector and beamline can be found in the initial proposal to the Jefferson Lab 
Program Advisory Committee (PAC) 30~\cite{pac30} and a subsequent PAC 36
update~\cite{pac36}.  While the currently-approved 120 days of beam time 
with the baseline detector configuration will allow \gx~an unprecedented opportunity 
to search for exotic hybrid mesons, the existing baseline design is inadequate 
for studying mesons or baryons with strange quarks.  This proposal focuses on developing 
additional detector capability that will allow \gx~to identify kaons and operate at 
high intensity.  This functionality is essential in order for the \gx~experiment to pursue its 
primary goal of solidifying our experimental understanding of hybrids by identifying {\em patterns} 
of hybrid mesons, both isoscalar and isovector, exotic and non-exotic, that are embedded in 
the spectrum of conventional mesons.

\subsection{Theoretical context}
\label{sec:theory}

Our understanding of how gluonic excitations manifest themselves within QCD 
is maturing thanks to recent results from lattice QCD. This numerical approach 
to QCD considers the theory on a finite, discrete grid of points in a manner that 
would become exact if the lattice spacing were taken to zero and the spatial extent of the calculation, {\it i.e.,} the ``box size," 
was made large. In practice, rather fine spacings and large boxes are used so that 
the systematic effect of this approximation should be small. The main limitation of
these calculations at present is the poor scaling of the numerical algorithms 
with decreasing quark mass - in practice most contemporary calculations use a 
range of artificially heavy light quarks and attempt to observe a trend as the light 
quark mass is reduced toward the physical value. Trial calculations at the physical 
quark mass have begun and regular usage is anticipated within a few years.

The spectrum of eigenstates of QCD can be extracted from correlation functions 
of the type $\langle 0 | \mathcal{O}_f(t) \mathcal{O}_i^\dag(0) | 0 \rangle$, where 
the $\mathcal{O}^\dag$ are composite QCD operators capable of interpolating a 
meson or baryon state from the vacuum. The time-evolution of the Euclidean 
correlator indicates the mass spectrum ($e^{-m_\mathfrak{n} t}$) and information 
about quark-gluon substructure can be inferred from matrix-elements 
$\langle \mathfrak{n} | \mathcal{O}^\dag |0 \rangle$. In a series of recent 
papers~\cite{Dudek:2009qf,Dudek:2010wm,Dudek:2011tt,Edwards:2011jj}, 
the Hadron Spectrum Collaboration has explored the spectrum of mesons and 
baryons using a large basis of composite QCD interpolating fields, extracting a 
spectrum of states of determined $J^{P(C)}$, including states of high internal excitation.

\begin{figure*}
\begin{center}
\includegraphics[width=\linewidth]{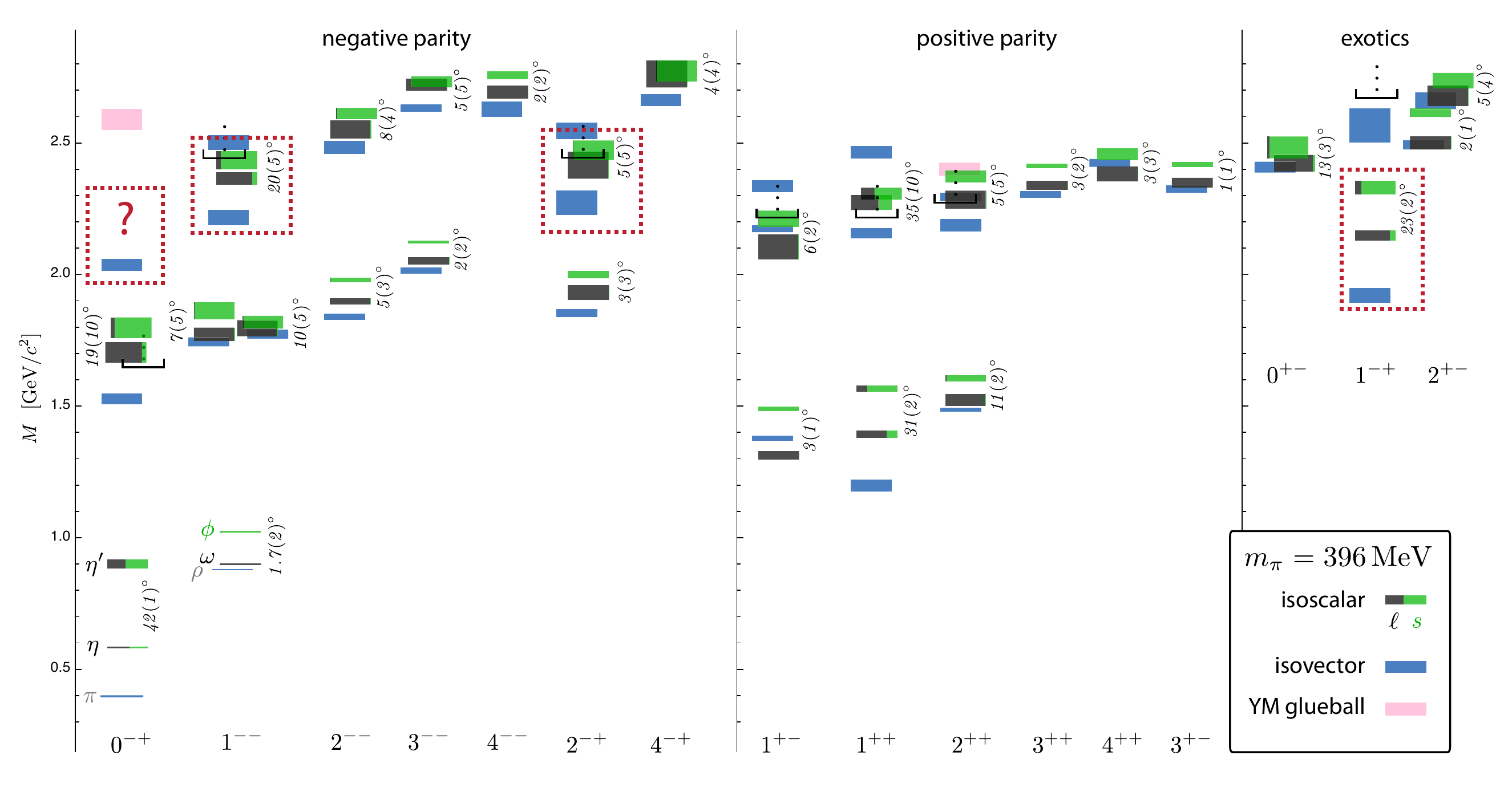}
\caption{\label{fig:lqcd_meson}A compilation of recent lattice QCD computations for both the
isoscalar and isovector light mesons from Ref.~\cite{Dudek:2011bn}, including 
$\ell\bar{\ell}$ $\left(|\ell\bar{\ell}\rangle\equiv (|u\bar{u}\rangle+|d\bar{d}\rangle)/\sqrt{2}\right)$ and
$s\bar{s}$ mixing angles (indicated in degrees).  The dynamical computation is
carried out with two flavors of quarks, light ($\ell$) and strange ($s$).  The $s$ quark
mass parameter is tuned to match physical $s\bar{s}$ masses, while the light quark mass parameters are
heavier, giving a pion mass of 396~MeV.  The black brackets with upward ellipses represent regions of the spectrum
where present techniques make it difficult to extract additional states.  
The dotted boxes indicate states that are interpreted as the lightest
hybrid multiplet -- the extraction of clear $0^{-+}$ states in this region is difficult in practice.}
\end{center}
\end{figure*}

As shown in Fig.~\ref{fig:lqcd_meson}, these calculations, for the first time, show a 
clear and detailed spectrum of exotic 
$J^{PC}$ mesons, with a lightest $1^{-+}$ lying a few hundred MeV below a $0^{+-}$ 
and two $2^{+-}$ states. Beyond this, through analysis of the matrix elements 
$\langle \mathfrak{n} | \mathcal{O}^\dag |0 \rangle$ for a range of different quark-gluon 
constructions, $\mathcal{O}$, we can infer \cite{Dudek:2011bn} that although the bulk of the 
non-exotic $J^{PC}$ spectrum has the expected systematics of a $q\bar{q}$ bound 
state system, some states are only interpolated strongly by operators featuring non-trivial 
gluonic constructions. One may interpret these states as non-exotic hybrid mesons, and, by combining them with
the spectrum of exotics, it is then possible to isolate a lightest hybrid supermultiplet of $(0,1,2)^{-+}$ and $1^{--}$ states, 
roughly 1.3 GeV heavier than the $\rho$ meson. The form of the operator that has strongest overlap 
onto these states has an $S$-wave $q\bar{q}$ pair in a color octet configuration and 
an exotic gluonic field in a color octet with $J_g^{P_gC_g}=1^{+-}$, a \emph{chromomagnetic} 
configuration. The heavier $(0,2)^{+-}$ states, along with some positive parity 
non-exotic states, appear to correspond to a $P$-wave coupling of the $q\bar{q}$ pair 
to the same chromomagnetic gluonic excitation.

A similar calculation for isoscalar states uses both $u\bar{u} + d\bar{d}$ and $s\bar{s}$ 
constructions and is able to extract both the spectrum of states and also their hidden 
flavor mixing. (See Fig.~\ref{fig:lqcd_meson}.)  The basic experimental pattern of significant 
mixing in $0^{-+}$ and $1^{++}$ 
channels and small mixing elsewhere is reproduced, and, for the first time, we are able to say 
something about the degree of mixing for exotic-$J^{PC}$ states.  In order to
probe this mixing experimentally, it is essential to be able to reconstruct decays to both strange and non-strange
final state hadrons.

\begin{figure*}
\begin{center}
\includegraphics[width=0.8\linewidth]{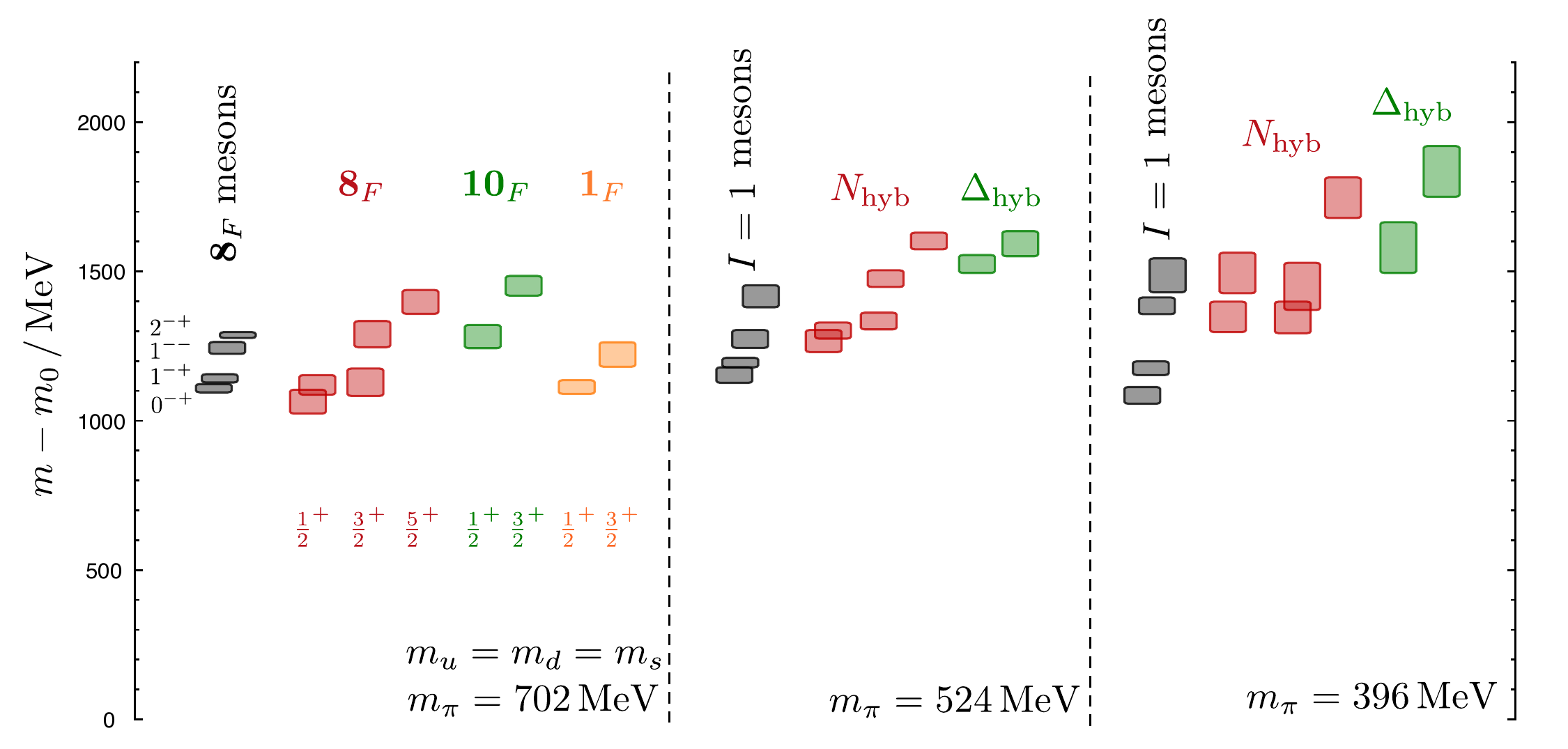}
\caption{\label{fig:baryon_meson} Spectrum of gluonic excitations in hybrid mesons (grey) and hybrid baryons (red, green, and orange) for three light quark masses.  The mass scale is $m-m_\rho$ for mesons and $m-m_N$ for baryons to approximately subtract the effect of differing numbers of quarks.  The left calculation is performed with perfect $SU(3)$-flavor symmetry, and hybrid members of the flavor octets ($8_F$), decuplet ($10_F$), and singlet ($1_F$) are shown.  The middle and right calculations are performed with a physical $s\bar{s}$ mass and two different values of $m_\pi$.}
\end{center}
\end{figure*}

A chromomagnetic gluonic excitation can also play a role in the spectrum of 
baryons:  constructions beyond the simple $qqq$ picture can occur when three quarks 
are placed in a color octet coupled to the chromomagnetic excitation. The 
baryon sector offers no ``smoking gun" signature for hybrid states, as all $J^P$ can be 
accessed by three quarks alone, but lattice calculations \cite{Edwards:2011jj} indicate that 
there are ``excess" nucleons with $J^P=1/2^+,~3/2^+,~5/2^+$ and excess $\Delta$'s 
with $J^P=1/2^+, 3/2^+$ that have a hybrid component. An interesting observation 
that follows from this study is that there appears to be a common energy cost for 
the chromomagnetic excitation, regardless of whether it is in a meson or baryon. 
In Fig.~\ref{fig:baryon_meson} we show the hybrid meson spectrum alongside the hybrid baryon 
spectrum with the quark mass contribution subtracted 
(approximately, by subtracting the $\rho$ mass from the mesons, and the nucleon mass from the baryons). 
We see that there appears to be a common scale $\sim 1.3$~GeV for the gluonic excitation, which does not 
vary significantly with varying quark mass.

Hybrid baryons will be challenging to extract experimentally because they lack ``exotic" character,
and can only manifest themselves by overpopulating the predicted spectrum with respect 
to a particular model. The current experimental situation of nucleon and $\Delta$~excitations is, 
however, quite contrary to the findings in the meson sector. Fewer baryon resonances are observed 
than are expected from models using three symmetric quark degrees of freedom, which does not 
encourage adding additional gluonic degrees of freedom. The current experimental efforts at 
Jefferson Lab aim to identify the relevant degrees of freedom which give rise to nucleon excitations.

While lattice calculations have made great progress at predicting the $N^\ast$~and $\Delta$~spectrum, 
including hybrid baryons~\cite{Edwards:2011jj,Dudek:2012ag}, calculations are also underway for 
$\Xi$~and $\Omega$~resonances. The properties of these multi-strange states are poorly known; only 
the $J^P$ of the $\Xi(1820)$ have been (directly) determined experimentally~\cite{Nakamura:2010zzi}. 
Previous experiments searching for Cascades were limited by low statistics and poor detector 
acceptance, making the interpretation of available data difficult. An experimental program on 
Cascade physics using the \gx~detector provides a new window of opportunity in hadron 
spectroscopy and serves as a complementary approach to the challenging study of broad and 
overlapping $N^\ast$~states. Furthermore, multi-strange baryons provide an important missing 
link between the light-flavor and the heavy-flavor baryons. Just recently many new beautiful 
baryons, $\Sigma_b^-$, $\Xi_b^-$, $\Omega_b^-$, have been discovered at the 
Tevatron~\cite{arXiv:0706.3868,arXiv:0706.1690, Aaltonen:2011wd,arXiv:0808.4142,arXiv:0905.3123}. 
We discuss $\Xi$~baryons further in Section~\ref{sec:xi_baryon}.

\subsection{Experimental context}
\label{sec:expcontext}

\gx~is ideally positioned to conduct a search for light quark exotics and provide complementary
data on the spectrum of light quark mesons.  It is anticipated that between now and
the time \gx~begins data taking, many results on the light quark spectrum will have
emerged from the BESIII experiment, which is currently attempting to collect about $10^8$ to $10^9$
$J/\psi$ and $\psi'$ decays.  These charmonium states decay primarily through $c\bar{c}$
annihilation and subsequent  hadronization into light mesons, making them an ideal
place to study the spectrum of light mesons.  In fact several new states have already been 
reported by the BESIII collaboration such as the $X(1835)$, $X(2120)$, and $X(2370)$ in 
$J/\psi\to\gamma X$, $X\to\eta'\pi\pi$~\cite{Ablikim:2010au}.  No quantum number assignment
for these states has been made yet, so it is not yet clear where they fit into the meson spectrum.
\gx~can provide independent confirmation of the existence of these states in a completely different
production mode, in addition to measuring (or confirming) their $J^{PC}$ quantum numbers.
This will be essential for establishing the location of these states in the meson spectrum.
The BESIII experiment has the ability to reconstruct virtually any combination of final
state hadrons, and, due to the well-known initial state, kinematic fitting can be
used to virtually eliminate background. The list of putative new states and, therefore, the list of channels 
to explore with \gx, is only expected to grow over the next few years as BESIII acquires and 
analyzes its large samples of charmonium data.

While the glue-rich $c\bar{c}$ decays of charmonium have long been hypothesized as
the ideal place to look for glueballs, decays of charmonium have also recently been used
to search for exotics.  The CLEO-c collaboration studied both $\pi^+\pi^-$ and
$\eta'\pi^\pm$ resonances in the decays of $\chi_{c1}\to\eta'\pi^+\pi^-$ and observed a
significant signal for an exotic $1^{-+}$ amplitude in the $\eta'\pi^\pm$ system~\cite{Adams:2011sq}.
The observation is consistent with the $\pi_1(1600)$ previously reported by E852 in
the $\eta'\pi$ system~\cite{Ivanov:2001rv}.  However, unlike E852, the CLEO-c analysis was 
unable to perform a model-independent extraction of the $\eta'\pi$ scattering amplitude
and phase to validate the resonant nature of the $1^{-+}$ amplitude.  A similar analysis
of $\chi_{c1}$ decays will most likely be performed by BESIII; however, even with an
order of magnitude more data, the final $\eta'\pi^+\pi^-$ sample is expected to be
just tens of thousands of events, significantly less than the proposed samples that will
be collected with \gx.  With the exception of this recent result from CLEO-c, the picture
in the light quark exotic sector, and the justification for building \gx, 
remains largely the same as it did at the time of the
original \gx~proposal; see Ref.~\cite{Meyer:2010ku} for a review.  All exotic candidates reported
to date are isovector $1^{-+}$ states ($\pi_1$).  With the addition of 
kaon identification, \gx~is capable of exploring all possible decay modes (non-strange
and strange) in order to establish not just one exotic state,
but a {\em pattern} of hybrid states with both exotic and non-exotic quantum numbers.

The idea that hybrids should also appear as supernumerary states in the spectrum of
non-exotic $J^{PC}$ mesons suggests an interesting interpretation of recent data in charmonium.
Three independent experiments have observed a state denoted $Y(4260)$~\cite{Aubert:2005rm,Coan:2006rv,He:2006kg,Yuan:2007sj};
it has $1^{--}$ quantum numbers but has no clear assignment in the arguably 
well-understood spectrum of $c\bar{c}$.  Even though the state is above $D\bar{D}$ 
threshold, it does not decay strongly to $D\bar{D}$ as the other $1^{--}$ $c\bar{c}$ 
states in that region do.  Its mass is about 1.2~GeV above the ground state $J/\psi$,
which is similar to the splitting observed in lattice calculations of light mesons and baryons.
If this state is a non-exotic hybrid, an obvious, but very challenging, experimental goal would be to
identify the exotic $1^{-+}$ $c\bar{c}$ hybrid member of the same multiplet, which should have 
approximately the same mass\footnote{Like the light quark mesons discussed in Sec.~\ref{sec:theory}, 
the expectation in charmonium is that a $1^{--}$ non-exotic hybrid would exist with about the same mass
as the $1^{-+}$ exotic charmonium hybrid~\cite{Dudek:2008sz,Liu:2012ze}.}.  It is not clear how to produce such a state with existing experiments.  
In the light quark sector, some have suggested that the recently discovered $Y(2175)$~\cite{Aubert:2006bu,Ablikim:2007ab,Shen:2009zze} 
is the strangeonium ($s\bar{s}$) 
analogue of the $Y(4260)$.  If this is true, \gx~is well-positioned to study this state and
search for its exotic counterpart.  We discuss this further in Section~\ref{sec:normalss}.

Recent CLAS results~\cite{Price:2004xm,Guo:2007dw} also suggest many opportunities to make 
advances in baryon spectroscopy.  
The CLAS collaboration investigated Cascade photoproduction in the reactions $\gamma p\to K^+K^+\,(X)$ as well as 
$\gamma p\to K^+K^+\pi^-\,(X)$ and among other things, determined the mass splitting of the 
ground state $(\Xi^-,\,\Xi^0)$ doublet to be $5.4\pm 1.8$~MeV/$c^2$, which is consistent with 
previous measurements. Moreover, the differential (total) cross sections for the $\Xi^-$ have been 
determined in the photon energy range from 2.75 to 3.85~GeV~\cite{Guo:2007dw}. 
The cross section results are consistent with a production mechanism of $Y^\ast\to \Xi^-K^+$ 
through a $t$-channel process. The reaction $\gamma p\to K^+K^+\pi^-\,[\Xi^0]$ was also studied 
in search of excited Cascade resonances, but no significant signal for an excited Cascade state, 
other than the $\Xi^-(1530)$, was observed. The absence of higher-mass signals is very likely 
due to the low photon energies and the limited acceptance of the CLAS detector. Equipped with 
a kaon identification system, the \gx~experiment will be well-suited to search for and study 
excited $\Xi$~resonances.

\section{Status of the \gx~experiment}


In the following section, we discuss the current status of the development of the 
baseline \gx~experiment.  The \gx~experiment was first presented to PAC~30 
in 2006~\cite{pac30}.   While beam time was not awarded for 12~GeV proposals 
at that PAC, the proposal suggested a three phase startup for \gx, which spanned 
approximately the first two calendar years of operation.  Phase I covered detector 
commissioning.  Phases II and III proposed a total of $7.5\times10^6$~s of 
detector live time at a flux of $10^7~\gamma$/s for physics commissioning and 
initial exploratory searches for hybrid mesons.  In 2010, an update of the experiment 
was presented to PAC~36 and a total of 120 days of beam time was granted for Phases I-III.  

In 2008, two critical detector components were ``de-scoped" from 
the design due to budgetary restrictions.  First, and most importantly, the forward 
Cherenkov particle identification system was removed.  The other component 
that was taken out was the level-three software trigger, which is needed for operating at a
photon flux greater than $10^7~\gamma$/s.  These changes severely impact the 
ultimate scientific goals and discovery potential of the \gx~experiment, as was 
noted in the PAC report:

\begin{quote}
Finally, the PAC would like to express its hope that the de-scoped Cherenkov 
detector be revisited at some time in the future. The loss of kaon identification 
from the current design is a real shame, but entirely understandable given the 
inescapable limitations on manpower, resources, and time.
\end{quote}

We now propose the development of a kaon identification system to be used 
during a high intensity ($>10^7~\gamma$/s) Phase~IV running of \gx.  Such a 
system and beam intensity will allow systematic exploration of higher-mass 
$s\bar{s}$ states, with the goal of identifying $s\bar{s}$ members of the hybrid 
nonets and studying $s\bar{s}$ and $\ell\bar{\ell}$ 
$\left( |\ell\bar{\ell}\rangle\equiv (|u\bar{u}\rangle+|d\bar{d}\rangle)/\sqrt{2}\right)$ mixing 
amongst isoscalar mesons.  A kaon identification system will also provide the
capability needed to study doubly-strange $\Xi$ baryons.  In the remainder of 
this section, we review our progress toward the baseline \gx~construction, 
which is aimed at Phases~I-III of running. Sections~\ref{sec:ss_meson} 
and~\ref{sec:xi_baryon} discuss the physics motivation for Phase~IV, and 
Section~\ref{sec:hardware} covers the Phase~IV hardware and beam time 
requirements.


\subsection{\gx~construction progress}

\begin{figure*}
\begin{center}
\includegraphics[width=0.8\linewidth]{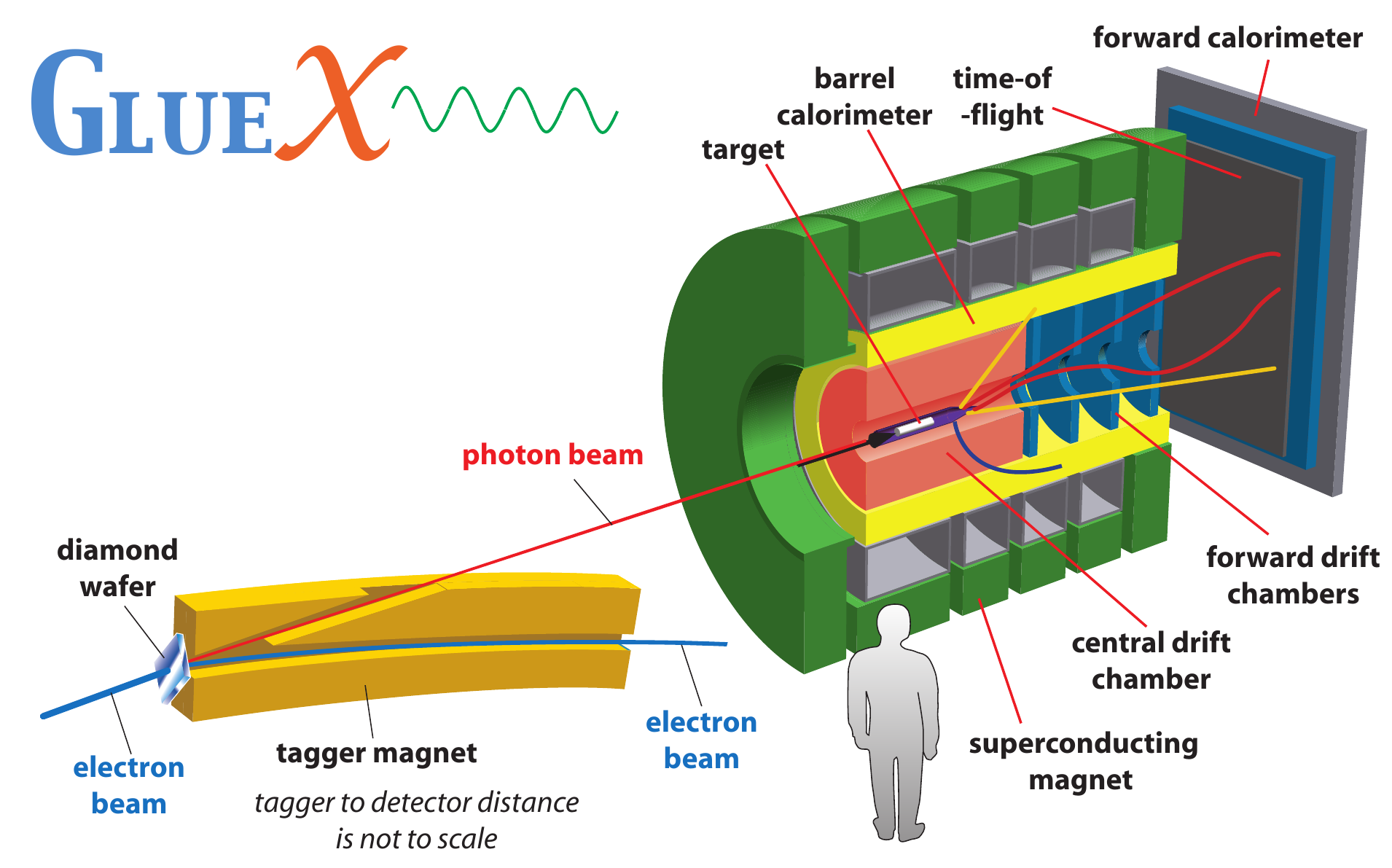}
\caption{\label{fig:detector}A schematic of the \gx~detector and beam.}
\end{center}
\end{figure*}

A schematic view of the \gx~detector is shown in Fig.~\ref{fig:detector}.  All major components of the detector are under
construction at Jefferson Lab or various collaborating institutions.  Beam for the experiment is derived from 
coherent bremsstrahlung radiation from a thin diamond wafer and delivered to a liquid hydrogen target.  The 
solenoidal detector has both central and forward tracking chambers as well as central and forward calorimeters.  
Timing and triggering are aided by a forward time of flight wall and a thin scintillator start counter that encloses the target.  We
briefly review the capabilities and construction status of each of the detector components below.  The civil construction of
Hall D is complete, and assembly of the detector within the hall is currently underway.  Table~\ref{tab:institutional_responsibilities}
briefly lists all of the \gx~collaborating institutions and their primary responsibilities.  The collaboration has grown
significantly in the past four years; newer members are noted in the table.

\begin{table*}
\begin{center}
\caption{\label{tab:institutional_responsibilities}A summary of \gx~institutions and
their responsibilities. The star ($\star$) indicates that the group has joined \gx~after 2008.}
\begin{tabular}{l|l}\hline\hline
Institution & Responsibilities \\ \hline
Arizona State U.$^{\star}$ & beamline polarimetry, beamline support \\
Athens & BCAL and FCAL calibration \\
Carnegie Mellon U. & CDC, offline software, management \\
Catholic U. of America & tagger system \\
Christopher Newport U. & trigger system \\
U. of Connecticut & tagger microscope, diamond targets, offline software\\
Florida International U. & start counter \\
Florida State U. & TOF system, offline software \\
U. of Glasgow$^{\star}$ & goniometer, beamline support \\
Indiana U. & FCAL, offline software, management \\
Jefferson Lab & FDC, data acquisition, electronics, infrastructure, management \\
U. of Massachusetts $^{\star}$ & target, electronics testing \\
Massachusetts Institute of Technology$^{\star}$ & forward PID, offline software \\
MEPHI$^{\star}$ & offline and online software \\ 
U. of North Carolina A\&T State$^{\star}$ & beamline support \\
U. of North Carolina, Wilmington$^{\star}$ & pair spectrometer \\
U. T\'ecnica Federico Santa Mar\'ia & BCAL readout \\
U. of Regina & BCAL, SiPM testing \\ \hline\hline
\end{tabular}
\end{center}
\end{table*}

\subsubsection{Beamline and Tagger}

The \gx~photon beam originates from coherent bremsstrahlung radiation produced by the 12~GeV electron beam
impinging on a $20~\mu m$ diamond wafer.  Orientation of the diamond and downstream collimation
produces a photon beam peaked in energy around 9~GeV with about 40\% linear polarization.
A coarse tagger tags a broad range of electron energy, while precision tagging in the coherent peak 
is performed by a tagger microscope.  A downstream pair spectrometer is utilized to measure photon conversions
and determine the beam flux.  Construction of the full system is underway.

Substantial work has also been done by the Connecticut group to
fabricate and characterize thin diamond radiators for \gx. This has included 
collaborations with the Cornell High Energy Synchrotron Source as well as industrial partners. 
Successful fabrication of 20~$\mu$m diamond radiators for \gx~seems possible.
The design of the goniometer system to manipulate the diamond has been completed by the Glasgow group and
is expected to be fabricated in 2012 by private industry. 

The tagger magnet and vacuum vessel are currently being manufactured by industrial contractors, 
with winding of the coil starting in late April of 2012. These elements are expected to be installed in 
the tagger hall over the next eighteen months. The design for the precision tagger ``microscope" 
has been developed at Connecticut, including the custom electronics for silicon photomultiplier (SiPM) readout.  
Beam tests of prototypes have been conducted, and construction
of the final system will begin in the fall of 2012.  The coarse tagger, which covers the entire energy range 
up to nearly the endpoint, will be built by the Catholic University group.  Work is underway to finalize the choice 
of photomultiplier tubes, and construction will start in the fall of 2012.

The groups from the University of North Carolina at Wilmington, North Carolina A\&T State, and Jefferson Lab are 
collaborating to develop and construct the pair spectrometer.  A magnet was obtained
from Brookhaven and modified to make it suitable for use in Hall D.
In addition, the Arizona State and Glasgow groups are collaborating to develop a technique
for accurately measuring the linear polarization of the beam.  Tests are planned in Mainz for the next year.

\subsubsection{Solenoid}

At the heart of the \gx~detector is the 2.2~T superconducting solenoid, which provides the essential 
magnetic field for tracking.  The solenoidal geometry also has the benefit of reducing electromagnetic
backgrounds in the detectors since low energy $e^+e^-$ pairs spiral within a small radius of the beamline.
The field is provided by four superconducting coils.  Three of the four have been tested up to the nominal
current of 1500~A -- the remaining coil was only tested to 1200~A due to a problem with power leads that
was unrelated to the coil itself.  No serious problems have been found, and the magnet has now been fully
assembled in Hall~D.  Both the cryogenic supply box and control systems will be developed this summer, and
the plan is to start power tests of the assembled magnet in January.

Two years ago, there was great concern about the operational capabilities of the magnet, and
 the DOE required that we begin the process of building a replacement magnet.  
The conceptual design was subcontracted to MIT.  The initial plan was to procure 
the technical design of the magnet with an option to build.  However, given the recent successful tests of
all the coils and the additional pressure to reduce overall project costs, development of the
replacement solenoid has been halted and will likely be cancelled unless there are significant problems
with the original magnet.

\subsubsection{Tracking}

Charged particle tracking is performed by two systems:  a central straw-tube drift chamber (CDC) and four six-plane
forward drift chamber (FDC) packages.  The CDC is composed of 28 layers of 1.5~m-long straw tubes.  The chamber
provides $r-\phi$ measurements for charged tracks.  Sixteen of the 28 layers have a $6^\circ$ stereo angle to 
supply $z$ measurements.  Each FDC package is composed of six planes of anode wires.  The cathode strips
on either side of the anode cross at $\pm75^\circ$ angles, providing a two-dimensional intersection 
point on each plane.  Design and construction of the CDC~\cite{VanHaarlem:2010yq} and FDC is 
ongoing at Carnegie Mellon University (CMU) and Jefferson Lab, respectively.  
The position resolution of the CDC and FDC is about $150~\mu$m and $200~\mu$m, respectively.  Together the
approximate momentum resolution is 2\%, averaged over the kinematical regions of interest.

Construction on the CDC began in May of 2010 with initial procurement and quality assurance of components and
the construction of a 400~ft$^2$ class 2000 cleanroom at CMU.
In August, the end plates were mounted on the inner shell and then
aligned. The empty frame was then moved into a vertical position for the installation of
the 3522 straw tubes. This work started in November of 2010 and continued until October of 2011, 
when the outer shell was installed on the chamber.  Stringing of the wires was completed in 
February of 2012, and all tension voltage and continuity checks were completed in March of 2012.
In May of 2012, the upstream gas plenum will be installed
on the chamber and the last labor-intensive phase started. This phase involves connecting each
of the exposed crimp pins on the upstream end plate to the correct channel on a transition
electronics board mounted on the end of the chamber. These boards have connectors that
hold the high-voltage and preamplifier cards.  The CDC will be delivered on time to
Jefferson Lab in April of 2013 for installation into the \gx~detector.

After successful studies with a full-scale prototype, the FDC construction started in the beginning of 2011,
with the entire production process carried out by Jefferson Lab in an off-site, 2000~ft$^2$ class 10,000 clean room. 
At the beginning of 2012, 
two months ahead of the schedule, two of the four packages had already been assembled.
Tests of the first package have started with cosmic rays in a system that
uses external chambers for tracking, scintillators for triggering,
and a DAQ system. The package has exhibited very good HV stability; however, low chamber efficiency was
traced to a problem with oxygen contamination.  This problem was mitigated by redesigning the gas seal between
the wire frames, and the revised design works as anticipated.  The repair will need to be performed on the chambers
that have already been fabricated; however, even with this delay, the production of all the FDC packages will be 
finished on schedule, by the end of this year.

\subsubsection{Calorimetry}

Like tracking, the \gx~calorimetry system consists of two detectors:  a barrel calorimeter with a 
cylindrical geometry (BCAL) and a forward lead-glass calorimeter with a planar geometry (FCAL).  
The primary goal of these systems is to detect photons that can be used to reconstruct $\pi^0$'s and $\eta$'s, 
which are produced in the decays of heavier states.  The BCAL is a relatively high-resolution 
sampling calorimeter, based on 1~mm double-clad Kuraray scintillating fibers embedded in a lead matrix.
It is composed of 48 4-m long modules; each module having a radial thickness of 15.1 radiation lengths.
Modules are read out on each end by silicon SiPMs, which are not adversely 
affected by the high magnetic field in the proximity of the \gx~solenoid flux return.
The forward calorimeter is composed of 2800 lead glass modules, stacked in a circular array.  Each
bar is coupled to a conventional phototube.  The fractional energy resolution of the combined calorimetry
system $\delta(E)/E$ is approximately $5\%$-$6\%/\sqrt{E~[\mathrm{GeV}]}$.  Monitoring systems for both
detectors have been designed by the group from the University of Athens.

All 48 BCAL calorimeter modules and a spare have been fabricated by the University of Regina
and are currently in storage at Jefferson Lab, waiting to be fitted with light guides and sensors.  
A full module was tested extensively in Hall B in 2006~\cite{Leverington:2008zz}; 
however, the readout for these tests used conventional phototubes instead of the
final SiPMs.  The production SiPMs are Hamamatsu S12045(X) MPPC arrays -- currently about half of them have
been delivered.  Testing of the SiPMs is  ongoing at Jefferson Lab and Universidad 
T\'ecnica Federico Santa Mar\'ia (USM), with good consistency between
each other and test data provided by Hamamatsu.  
The production light guides are being machined and polished at the USM. 
The first-article samples have been received and were used to instrument a final prototype, which 
is currently (in spring 2012) being tested with electrons in Hall B.  
Installation of the completed modules in the hall is expected to begin in the spring of 2013.

The 2800 lead glass modules needed for the FCAL have been assembled at Indiana University and shipped 
to Jefferson Lab.  All of the PMTs will be powered by custom-built Cockroft-Walton style photomultiplier
bases~\cite{Brunner:1998fh} in order to reduce cable mass, power dissipation, and high voltage control
system costs.  The design, fabrication, and testing of the bases is underway at Indiana University.
All components for the FCAL will be delivered to Jefferson Lab for assembly to begin in the fall of 2012.  A
25-block array utilizing the final design of all FCAL components was constructed and tested with
electrons in Hall~B in the spring of 2012; preliminary results indicate that the performance meets or
exceeds expectations.

\subsubsection{Particle ID and timing}

The existing particle ID capabilities are derived from several subsystems.  A dedicated 
forward time-of-flight wall (TOF), which is constructed from two planes of 2.5~cm-thick 
scintillator bars, provides about 70~ps timing resolution on forward-going tracks within 
about $10^\circ$ of the beam axis.  This information is complemented by time-of-flight data 
from the BCAL and specific ionization ($dE/dx$) measured with the CDC, both of which are 
particularly important for identifying the recoil proton in $\gamma p\to Xp$ reactions.  Finally,
identification of the beam bunch, which is critical for timing measurements, is performed by
a thin start counter that surrounds the target.

For the TOF system, the type of plastic scintillator and photomultiplier tube has been chosen. We have built a 
prototype with these choices that achieves 100 ps resolution for mean time from a single 
counter read-out from both ends. The system consists of two planes of such counters, implying
that the demonstrated two-plane resolution is 70 ps.  The photomultiplier purchase is underway. The light guide 
design and material choice have been finalized. We are working on the design of the 
mechanical support and have a working concept. Soon we plan to build a full-scale mock-up 
of the central counters around the beam line to prove the mechanical design.

Engineering drawings for the start counter are under development.  The counters and the 
electronics have to fit into a narrow space between the target vacuum chamber and the 
inner wall of the Central Drift Chamber.  Prototypes have obtained 
time resolution of 300 to 600~ps, depending on the position 
of the hit along the length of the counter. The final segmentation has been fixed. 
SiPMs will be used for readout because they can be placed in the high 
magnetic field environment very close to the counters, thereby
preserving scintillation light. The design of the 
SiPM electronics is about to start, and a final prototype of the scintillator assembly is under development.

The combined PID system in the baseline design is sufficient for identification of most
protons in the kinematic regions of interest for \gx.  The forward PID can be used
to enhance the purity of the charged pion sample.  However, the combined momentum, 
path length, and timing resolution only allows for exclusive kaon identification for
track momenta less than 2.0~GeV/$c$.

\subsection{Initial physics goals and sensitivity}

Phases~I-III of the \gx~physics program provide an excellent opportunity for both the study of 
conventional mesons and the search for exotic mesons in photoproduction.  Reconstructable final states will 
be limited to those decaying into non-strange states:  $\pi$, $\eta$, $\eta^\prime$, and $\omega$.
Table~\ref{tab:exotic_modes} summarizes the expected lowest mass exotics and possible decay modes.
Initial searches will likely focus on the $\pi_1$ isovector triplet and the $\eta_1$ isoscalar.  It will also be 
important to try to establish the other (non-exotic) members of the hybrid multiplet:  the $0^{-+}$, $1^{--}$, and $2^{-+}$
states.  Finally, the initial data may provide an opportunity to search for the heavier exotic $b_2$ and $h_2$ states.
Current lattice QCD predictions are that both the $1^{-+}$ ($\eta_1$ and $\eta_1'$) and $2^{+-}$ 
($h_2$ and $h_2'$) isoscalars exhibit relatively little $\ell\bar{\ell}-s\bar{s}$ mixing (Fig.~\ref{fig:lqcd_meson}).  
It will be impossible to test this prediction or, 
if it is correct, to identify the $\eta_1'$ and $h_2'$ states without further kaon identification hardware.

One reaction of interest is $\gamma p\to \pi^+\pi^-\pi^+ n$.  The $(3\pi)^\pm$ system has been 
studied extensively with data from E852~\cite{Adams:1998ff,Dzierba:2005jg}, COMPASS~\cite{Alekseev:2009aa}, 
and CLAS~\cite{Nozar:2008aa}, with
COMPASS reporting evidence for exotic $\pi_1(1600)\to\rho\pi$ decay.  
In an attempt to exercise our software framework, a full analysis of mock \gx~data 
was performed with this channel.  A \textsc{geant}-based simulation was developed to
model all active and inactive material within the detector volume.  Individual hits  and
background were simulated on all detector components, and reconstruction algorithms
were utilized to reconstruct each event without knowledge of the true generated particles.  Analysis
of any final state depends on first exclusively selecting the reaction of interest and then
performing an amplitude analysis to separate the various resonances that decay to that
stable final state.  

In order to test the detector capabilities for suppressing cross feed from other non-signal reactions, 
a \textsc{Pythia}-based generator was used to generate inclusive $\gamma{p}$ photoproduction at $E_\gamma = 9$~GeV.
The signal events ($\gamma p \to \pi^+\pi^-\pi^+n$) were generated at a level of about 2.5\% of 
the total hadronic cross section.
After optimizing all analysis criteria a signal selection efficiency of 25\% and a signal-to-background
ratio of 2:1 were achieved.  About 20\% of the total background originated from kaons misidentified as pions.
The other backgrounds included protons being misidentified as pions or extra $\pi^0$'s in the event
that went undetected.  This study, conducted in 2011, motivated a more detailed simulation of particle identification 
systems and tracking resolution along with enhancements in tracking efficiency.  This work is still
under development, and we expect that these enhanced algorithms along with improvements in 
analysis technique, such as kinematic fitting, will provide at least an order of magnitude further background 
suppression.  Reducing the background to the percent level is essential for enhancing sensitivity in the amplitude analysis.

\begin{figure*}
\begin{center}
\includegraphics[width=0.7\linewidth]{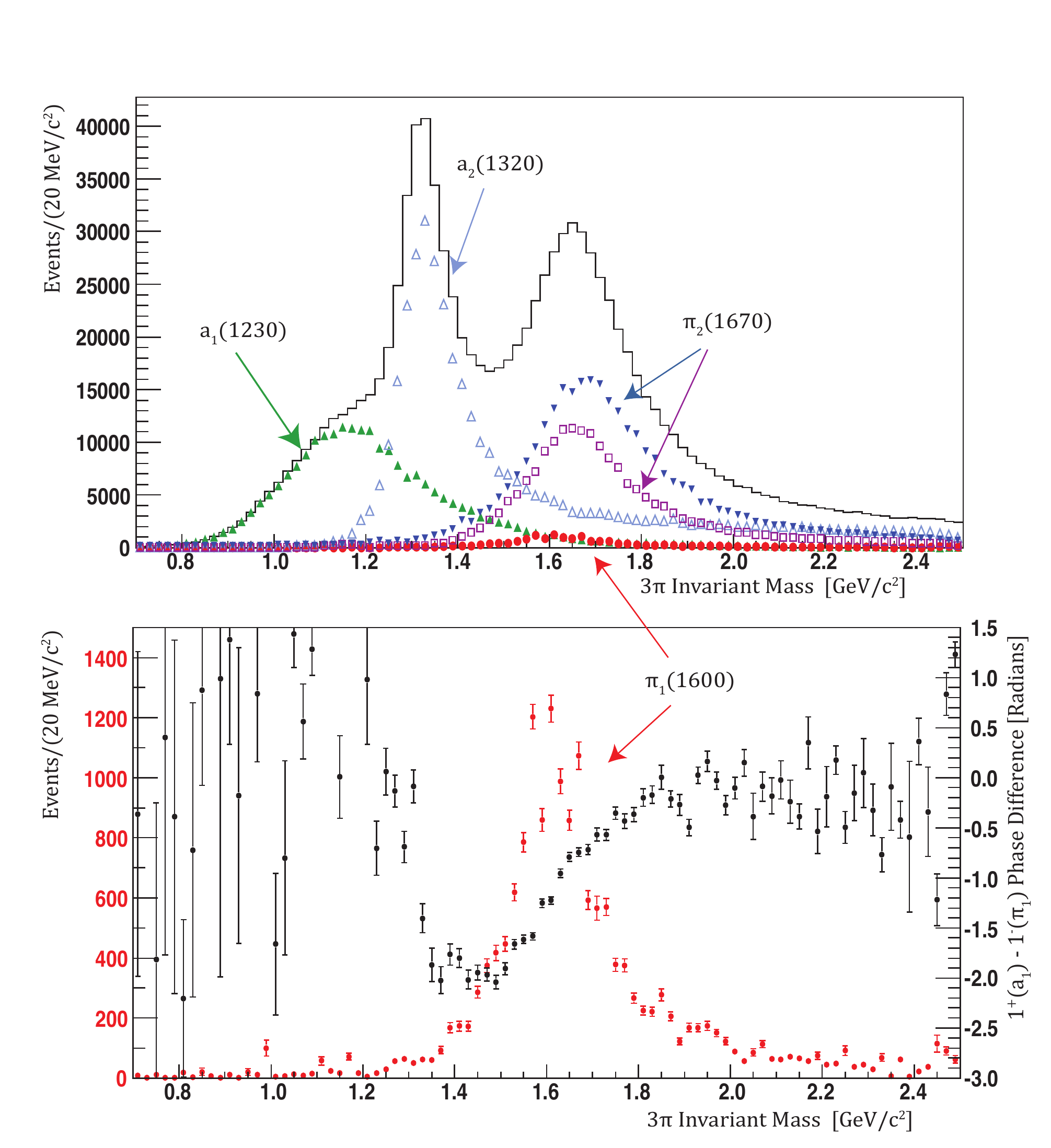}
\caption{\label{fig:amp_analysis}A sample amplitude analysis result for the $\gamma p \to \pi^+\pi^-\pi^+ n$ channel with
\gx.  (top) The invariant mass spectrum as a function of $M(\pi^+\pi^-\pi^+)$ is shown by the solid histogram.  The results
of the amplitude decomposition into resonant components in each bin is shown with points and error bars.  (bottom)  The
exotic amplitude, generated at a relative strength of 1.6\%, is cleanly extracted (red points).  The black points show the
phase between the $\pi_1$ and $a_1$ amplitudes.}
\end{center}
\end{figure*}

The sensitivity to small amplitudes that is provided by the \gx~detector acceptance and resolution was tested
by performing an amplitude analysis on a sample of purely generated $\gamma p\to \pi^+\pi^-\pi^+ n$ events 
that has been subjected to full detector simulation and reconstruction as discussed above.  Several conventional resonances, 
the $a_1$, $\pi_2$, and $a_2$, were 
generated along with a small ($<2\%$) component of exotic $\pi_1$.  The result of the fit is shown in Figure~\ref{fig:amp_analysis}.
This study indicates that with a pure sample of reconstructed decays, the \gx~detector provides excellent sensitivity
to rare exotic decays.  The analysis sensitivity will ultimately be limited by the ability to suppress and
parametrize backgrounds in the amplitude analysis that arise from improperly reconstructed events, as noted above.
While it is difficult to quantitatively predict the sensitivity with little knowledge of true detector performance or
actual physics backgrounds, increased statistics and improved kaon identification will certainly lead to
enhancements in sensitivity.  

In order to take full advantage of \gx~statistics, high sample purity, which only comes
with enhanced particle identification, is absolutely essential.  For comparison, 
the 2008 run of COMPASS is expected to yield a final analyzable sample of order $10^8$ events 
in the $\pi p\to3\pi p$ channel~\cite{Haas:2011rj}, statistics comparable to the approved \gx~Phase II and III running.
COMPASS will also have capability to positively identify kaons in the final state, enabling both high sample purity and
the ability to explore $s\bar{s}$ states.
While \gx~will utilize the complementary photoproduction mechanism, which has been argued to be rich in exotic 
production, it is important that both the statistical precision and sample purity remain competitive with 
state-of-the-art hadron beam experiments in order to maximize the collective and complementary physics output of
all experiments.

\section{\boldmath Study of $s\bar{s}$ Mesons}
\label{sec:ss_meson}

The primary goal of the \gx~experiment is to conduct a definitive mapping of states in the light meson sector with an
emphasis on searching for exotic mesons.  Ideally, we would like to produce the experimental analogue of the
lattice QCD spectrum pictured in Fig.~\ref{fig:lqcd_meson}, enabling a direct test of our understanding of 
gluonic excitations in QCD.  In order to achieve this, one must be able to reconstruct strange final states, 
as observing decay patterns of mesons has been one of the primary mechanisms of inferring quark flavor 
content.  An example of this can be seen by examining the two lightest isoscalar $2^{++}$ mesons in the lattice
QCD calculation in Fig.~\ref{fig:lqcd_meson}.  The two states are nearly pure flavors with only a small ($11^\circ$) mixing
in the $\ell\bar{\ell}$ and $s\bar{s}$ basis.  A natural experimental assignment for these two states are the $f_2(1270)$
and the $f_2'(1525)$.  An experimental study of decay patterns shows that 
$\mathcal{B}(f_2(1270)\to K K)/\mathcal{B}(f_2(1270)\to \pi\pi)\approx 0.05$ and 
$\mathcal{B}(f_2'(1525)\to \pi\pi)/\mathcal{B}(f_2'(1525)\to K K) \approx 0.009$~\cite{Nakamura:2010zzi}, which support the 
prediction of an $f_2(1270)$ ($f_2'(1525)$) with a dominant $\ell\bar{\ell}$ ($s\bar{s}$) component.  Without
the ability to identify final state kaons, \gx~is incapable of providing similarly valuable experimental data to aid
in the interpretation of the hybrid spectrum.

\subsection{\boldmath Exotic $s\bar{s}$ states}

While most experimental efforts to date have focused on the lightest isovector exotic meson, the 
$J^{PC}=1^{-+}$ $\pi_1(1600)$, lattice QCD clearly predicts a rich spectrum of both isovector and
isoscalar exotics, the latter of which may have mixed $\ell\bar{\ell}$ and $s\bar{s}$ flavor content.
A compilation of the ``ground state" exotic hybrids is listed in Table~\ref{tab:exotic_modes} along with
theoretical estimates for masses, widths, and key decay modes.  It is expected that initial searches 
with the baseline \gx~hardware will be targeted primarily at the $\pi_1$ state.  Searches for the $\eta_1$, $h_0$, and $b_2$ may be 
statistically challenging, depending on the masses of these states and the production cross sections.
With increased statistics and kaon identification, the search scope can be broadened to include these
heavier exotic states in addition to the $s\bar{s}$ states:  $\eta_1'$, $h_0'$, and $h_2'$.  The $\eta_1'$ and
$h_2'$ are particularly interesting because some models predict these states to be relatively narrow and
they should decay through well-established kaon resonances.

The observation of various $\pi_1$ states has been reported in the literature for over fifteen
years, with some analyses based on millions of events.  However, it is safe to say that there exists
a fair amount of skepticism regarding the assertion that unambiguous experimental evidence
exists for exotic hybrid mesons.  If the scope of exotic searches with \gx~is narrowed to only include
the lightest isovector $\pi_1$ state, the ability for \gx~to comprehensively address the question of the existence
of gluonic excitations in QCD is greatly diminished.  On the other hand, clearly identifying all exotic members
of the lightest hybrid multiplet, the three exotic $\pi_1^{\pm,0}$ states and the exotic $\eta_1$ and $\eta_1'$, 
which can only be done by systematically studying a large number of
strange and non-strange decay modes, would provide unambiguous experimental confirmation of exotic mesons.
A study of decays to kaon final states could demonstrate that the $\eta_1$ candidate is dominantly $\ell\bar{\ell}$
while the $\eta_1'$ candidate is $s\bar{s}$, as predicted by initial lattice QCD calculations.  Such a discovery
would represent a substantial improvement in the experimental understanding of exotics -- an improvement that
cannot be made by simply collecting more data with the baseline \gx~detector configuration.
In addition, further identification of members of the $0^{+-}$ and $2^{+-}$ nonets as well as measuring the
mass splittings with the $1^{+-}$ states will validate the phenomenological picture that is emerging from lattice QCD
of these states as $P$-wave couplings of a gluonic field with a color-octet $q\bar{q}$ system.

\begin{table*}\centering
\caption{\label{tab:exotic_modes}
A compilation of exotic quantum number hybrid approximate masses, widths, and decay predictions.
Masses are estimated from dynamical LQCD calculations with $M_\pi = 396~\mathrm{MeV}/c^2$~\cite{Dudek:2011bn}.  
The PSS (Page, Swanson and Szczepaniak) and IKP (Isgur, Kokoski and Paton) model widths are from Ref.~\cite{Page:1998gz},
with the IKP calculation based on the model in Ref.~\cite{Isgur:1985vy}.  The total widths have a mass 
dependence, and Ref.~\cite{Page:1998gz} uses somewhat different mass values than suggested by the most recent
lattice calculations~\cite{Dudek:2011bn}.
Those final states marked with a dagger ($\dagger$) are ideal for experimental exploration 
because there are relatively few stable particles in the final state or moderately narrow 
intermediate resonances that may reduce combinatoric background.  
(We consider $\eta$, $\eta^\prime$, and $\omega$ to be stable final state particles.)}
\begin{tabular}{ccccccc}\hline\hline
 &  Approximate & $J^{PC}$ & \multicolumn{2}{c}{Total Width (MeV)} & 
Relevant Decays & Final States \\ 
         & Mass (MeV) &  & PSS & IKP & &  \\ \hline
$\pi_{1}$   & 1900 & $1^{-+}$ &  $80-170$ & $120$ & 
$b_{1}\pi^\dagger$, $\rho\pi^\dagger$, $f_{1}\pi^\dagger$, $a_{1}\eta$, $\eta^\prime\pi^\dagger$ & $\omega\pi\pi^\dagger$, $3\pi^\dagger$, $5\pi$, $\eta 3\pi^\dagger$, $\eta^\prime\pi^\dagger$  \\
$\eta_{1}$  & 2100 & $1^{-+}$ &  $60-160$ & $110$ &
$a_{1}\pi$, $f_{1}\eta^\dagger$, $\pi(1300)\pi$ & $4\pi$, $\eta 4\pi$, $\eta\eta\pi\pi^\dagger$ \\ 
$\eta^{\prime}_{1}$ & 2300 & $1^{-+}$ &  $100-220$ & $170$ &
$K_{1}(1400)K^\dagger$, $K_{1}(1270)K^\dagger$, $K^{*}K^\dagger$ & $KK\pi\pi^\dagger$, $KK\pi^\dagger$, $KK\omega^\dagger$ \\ \hline
$b_{0}$     & 2400 & $0^{+-}$ & $250-430$ & $670$ &
$\pi(1300)\pi$, $h_{1}\pi$ & $4\pi$ \\
$h_{0}$     &  2400 & $0^{+-}$ & $60-260$  & $90$  &
$b_{1}\pi^\dagger$, $h_{1}\eta$, $K(1460)K$ & $\omega\pi\pi^\dagger$, $\eta3\pi$, $KK\pi\pi$ \\
$h^{\prime}_{0}$    & 2500& $0^{+-}$ & $260-490$ & $430$ &
$K(1460)K$, $K_{1}(1270)K^\dagger$, $h_{1}\eta$ & $KK\pi\pi^\dagger$, $\eta3\pi$ \\ \hline
$b_{2}$     & 2500 & $2^{+-}$ &    $10$ & $250$ &
$a_{2}\pi^\dagger$, $a_{1}\pi$, $h_{1}\pi$ & $4\pi$, $\eta\pi\pi^\dagger$ \\
$h_{2}$     & 2500 & $2^{+-}$ &    $10$ & $170$ &
$b_{1}\pi^\dagger$, $\rho\pi^\dagger$ & $\omega\pi\pi^\dagger$, $3\pi^\dagger$ \\
$h^{\prime}_{2}$  & 2600 & $2^{+-}$ &    $10-20$ &  $80$ &
$K_{1}(1400)K^\dagger$, $K_{1}(1270)K^\dagger$, $K^{*}_{2}K^\dagger$ & $KK\pi\pi^\dagger$, $KK\pi^\dagger$\\
\hline\hline
\end{tabular}
\end{table*}

\subsection{\boldmath Non-exotic $s\bar{s}$ mesons}
\label{sec:normalss}

As discussed in Section~\ref{sec:theory}, one expects the lowest-mass hybrid multiplet
to contain $(0,1,2)^{-+}$ states and a $1^{--}$ state that all have about the same mass
and correspond to an $S$-wave $q\bar{q}$ pair coupling to the gluonic
field in a $P$-wave.  For each $J^{PC}$ then we expect an isovector triplet and a pair of isoscalar
states in the spectrum.  Of the four sets of $J^{PC}$ values for the lightest hybrids, only the $1^{-+}$ is exotic.  
The other hybrid states will appear as supernumerary states in the spectrum of conventional mesons.
The ability to clearly identify these states depends on having a thorough and complete
understanding of the meson spectrum.  Like searching for exotics, a complete mapping
of the spectrum of non-exotic mesons requires the ability to systematically study many strange and non-strange
final states.  Other experiments, such as BESIII, are 
carefully studying this region utilizing of order $10^9$ hadronic decays of charmonium and have outstanding
capability to cleanly study any possible final state.  While the production mechanism of \gx~is
complementary to that of charmonium decay and is thought to enhance hybrid production, it
is essential that the detector capability and statistical precision of the data set be 
competitive with other contemporary experiments in order to maximize the collective
experimental knowledge of the meson spectrum.  Hybrid mesons with non-exotic $J^{PC}$ are
also expected to appear in the spectrum in the proximity of the exotic $(0,2)^{+-}$ states.

Given the recent developments in charmonium (briefly discussed in Section~\ref{sec:expcontext}),
a state that has attracted a lot of attention in the $s\bar{s}$ spectrum is the $Y(2175)$, which is
assumed to be an $s\bar{s}$ vector meson ($1^{--}$).  The
$Y(2175)$ has been observed to decay to $\pi\pi\phi$ and has been produced in both
$J/\psi$ decay~\cite{Ablikim:2007ab} and $e^+e^-$ collisions~\cite{Aubert:2006bu,Shen:2009zze}.  The state is a
proposed analogue of the $Y(4260)$ in charmonium, a state that is also about 1.2 GeV heavier than 
the ground state triplet ($J/\psi$)
and has a similar decay mode:  $Y(4260)\to\pi\pi J/\psi$.  The $Y(4260)$ has no obvious interpretation
in the charmonium spectrum and has been speculated to be a hybrid 
meson~\cite{Close:2005iz,Zhu:2005hp,Kou:2005gt,Luo:2005zg}, which, by loose analogy,
leads to the implication that the $Y(2175)$ might also be a hybrid candidate.  It should be noted
that the spectrum of $1^{--}$ $s\bar{s}$ mesons is not as well-defined experimentally as the $c\bar{c}$ system; 
therefore, it is not clear that the $Y(2175)$ is a supernumerary state.  However, \gx~is ideally suited
to study this system.  We know that vector mesons are copiously produced in photoproduction; therefore,
with the ability to identify kaons, a precision study of the $1^{--}$ $s\bar{s}$ spectrum can be conducted
with \gx.  Some have predicted~\cite{Ding:2007pc} that the potential hybrid nature of the $Y(2175)$ can be explored by
studying ratios of branching fractions into various kaonic final states.  
In addition, should \gx~be able to conclude that the $Y(2175)$ is in fact a supernumerary 
vector meson, then a search can be made for the exotic $1^{-+}$ $s\bar{s}$ member of the multiplet ($\eta_1'$), 
evidence of which would provide a definitive interpretation of the $Y(2175)$ and likely have implications
on how one interprets charmonium data.

\section{\boldmath $\Xi$ baryons}
\label{sec:xi_baryon}

The spectrum of multi-strange hyperons is poorly known, with only a
few well-established resonances. Among the
doubly-strange states, the two ground-state Cascades, the octet
member~$\Xi(1320)$ and the decuplet member $\Xi(1530)$, have four-star
status in the PDG~\cite{Nakamura:2010zzi}, with only four other
three-star candidates. On the other hand, more than 20~$N^\ast$ and
$\Delta^\ast$ resonances are rated with at least three stars in the
PDG. Of the six $\Xi$~states that have at least three-star ratings in
the PDG, only two are listed with weak experimental evidence for their
spin-parity $(J^P)$ quantum numbers: $\Xi(1530)\frac{3}{2}^+$~\cite{Aubert:2008ty},
$\Xi(1820)\frac{3}{2}^-$~\cite{Biagi:1986vs}. All other $J^P$~assignments 
are based on quark-model predictions. Flavor~$SU(3)$ symmetry predicts
as many $\Xi$~resonances as $N^\ast$ and $\Delta^\ast$~states
combined, suggesting that many more Cascade resonances remain
undiscovered. The three lightest quarks, $u$, $d$, and $s$, have 27
possible flavor combinations: $3\otimes 3\otimes 3 = 1\oplus 8\oplus
8\,^{\prime}\oplus 10$ and each multiplet is identified by its spin
and parity,~$J^P$. Flavor $SU(3)$ symmetry implies that the members of
the multiplets differ only in their quark makeup, and that the basic
properties of the baryons should be similar, although the symmetry is
known to be broken by the strange-light quark mass difference.
The octets consist of $N^*$, $\Lambda^*$, $\Sigma^*$, and
$\Xi^*$~states. We thus expect that for every $N^*$ state, there
should be a corresponding $\Xi^*$~state \emph{with similar
  properties}. Additionally, since the decuplets consist of
$\Delta^*$, $\Sigma^*$, $\Xi^*$, and  $\Omega^*$~states, we also
expect for every $\Delta^*$ state to find a decuplet~$\Xi^*$ with
similar properties.  

\subsection{\boldmath $\Xi$ spectrum and decays}

The $\Xi$~hyperons have the unique feature of double strangeness:
$|ssu\,\rangle$ and $|ssd\,\rangle$. If the confining potential is
independent of quark flavor, the energy of spatial excitations of a
given pair of quarks is inversely proportional to their reduced
mass. This means that the lightest excitations in each partial wave
are between the two strange quarks. In a spectator decay model, such
states will not decay to the ground state $\Xi$ and a pion because of
orthogonality of the spatial wave functions of the two strange quarks
in the excited state and the ground state. This removes the decay
channel with the largest phase space for the lightest states in each
partial wave, substantially reducing their widths. Typically,
$\Gamma_{\Xi^\ast}$ is about $10-20$~MeV for the known lower-mass
resonances, which is $5-30$~times narrower than for $N^\ast$,
$\Delta^\ast$, $\Lambda^\ast$, and $\Sigma^\ast$~states. These features, combined
with their isospin of 1/2, render possible a wide-ranging program on
the physics of the Cascade hyperon and its excited states. Until recently, the study
of these hyperons has centered on their production in
$K^- p$~reactions; some $\Xi^\ast$~states were found using high-energy
hyperon beams. Photoproduction appears to be a very promising 
alternative. Results from earlier kaon beam experiments indicate that
it is possible to produce the $\Xi$~ground state through the decay of
high-mass $Y^\ast$~states~\cite{Tripp:1967kj,Burgun:1969ee,Litchfield:1971ri}.
It is therefore possible to produce Cascade resonances through 
$t$-channel photoproduction of hyperon resonances using the
photoproduction reaction $\gamma p\to K
K\,\Xi^{(\ast)}$~\cite{Price:2004xm,Guo:2007dw}. 

In briefly summarizing a physics program on Cascades, it would be
interesting to see the lightest excited $\Xi^\ast$~states in certain
partial waves decoupling from the $\Xi\pi$~channel, confirming the
flavor independence of confinement. Measurements of the isospin
splittings in spatially excited Cascade states are also possible for
the first time in a spatially excited hadron. Currently, these
splittings like $n-p$ or $\Delta^0 - \Delta^{++}$ are only available
for the octet and decuplet ground states, but are hard to measure in
excited $N,~\Delta$ and $\Sigma,~\Sigma^\ast$~states, which are very
broad. The lightest Cascade baryons are expected to be narrower, and
measuring the $\Xi^- - \Xi^0$ splitting of spatially excited
$\Xi$~states remains a strong possibility. These measurements are an
interesting probe of excited hadron structure and would provide
important input for quark models, which describe the isospin splittings
by the $u$- and $d$-quark mass difference as well as by the
electromagnetic interactions between the quarks.

\subsection{\boldmath $\Xi$ searches using the \gx~experiment}

The Cascade octet ground states $(\Xi^0,\,\Xi^-)$ can be studied in
the \gx~experiment via exclusive $t$-channel (meson exchange)
processes in the reactions:
\begin{eqnarray}
\label{Equation:GroundOctet}
\gamma p\,\to\, K\,Y^\ast\,&\to&\,K^+\, (\,\Xi^-\,K^+\,), \nonumber\\
 &~& K^+\,(\,\Xi^0\,K^0\,), \nonumber\\
  &~& K^0\,(\,\Xi^0\,K^+\,)\,.
\end{eqnarray}
The production of such two-body systems involving a $\Xi$~particle
also allows for studying high-mass $\Lambda^\ast$ and $\Sigma^\ast$
states. The Cascade decuplet ground state, $\Xi(1530)$, and other
excited Cascades decaying to $\Xi\pi$ can be searched for and studied
in the reactions:
\begin{eqnarray}
\label{Equation:GroundDecuplet}
\gamma p\,\to\, K\,Y^\ast\,&\to&\,K^+\,(\,\Xi\,\pi\,)\,K^0,\nonumber\\
&~&K^+\,(\,\Xi\,\pi\,)\,K^+,\nonumber\\ 
&~& K^0\,(\,\Xi\,\pi\,)\,K^+\,.
\end{eqnarray}
The lightest
excited $\Xi$~states are expected to decouple from $\Xi\pi$ and can
be searched for and studied in their decays to $\Lambda \bar{K}$ and $\Sigma \bar{K}$:
\begin{eqnarray}
\label{Equation:Excited}
\gamma p\,\to\, K\,Y^\ast\,&\to&\,K^+\,(\,\bar{K}\Lambda\,)_{\Xi^{-\ast}}\,K^+,\nonumber\\
&~&K^+\,(\,\bar{K}\Lambda\,)_{\Xi^{0\ast}}\,K^0 ,\nonumber\\
&~&K^0\,(\,\bar{K}\Lambda\,)_{\Xi^{0\ast}}\,K^+,\\[1ex]
\gamma p\,\to\, K\,Y^\ast\,&\to&\,K^+\,(\,\bar{K}\Sigma\,)_{\Xi^{-\ast}}\,K^+,\nonumber\\
&~&K^+\,(\,\bar{K}\Sigma\,)_{\Xi^{0\ast}}\,K^0 ,\nonumber\\
&~&K^0\,(\,\bar{K}\Sigma\,)_{\Xi^{0\ast}}\,K^+.
\end{eqnarray}

Backgrounds from competing hadronic reactions will be reducible because 
of the unique signature provided by the two associated kaons in 
$\Xi$ photoproduction in combination with secondary vertices, 
stemming from weak decays of ground-state hyperons.  However, decays such as
$Y^\ast\to\phi\Lambda,~\phi\Sigma$ might contribute for certain final states. Larger
contributions to the background will more likely come from events with
pions misidentified as kaons as well as other reconstruction and detector
inefficiencies. To extract small Cascade signals at masses above the
$\Xi(1530)$, it will therefore be important to reduce the background
by reconstructing kinematically complete final states that include kaons.
A full exclusive reconstruction also enhances the possibility of being able
to measure the $J^P$ of these states.

\section{\gx~Hardware and Beam Time Requirements}
\label{sec:hardware}

In order to maximize the discovery capability of \gx, an enhanced capability to
identify kaons coupled with an increase in statistical precision is needed.
In this section, we detail those needs as well as provide two possible conceptual 
designs for a kaon detector.  To maximize sensitivity, we propose a gradual 
increase in the photon flux towards the \gx~design of $10^8~\gamma/$s in the peak of the coherent
bremsstrahlung spectrum ($8.4~\mathrm{GeV} < E_\gamma < 9.0~\mathrm{GeV}$).  
Yield estimates, assuming an average flux
of $5\times10^7~\gamma/$s, are presented.  In order to minimize the bandwidth
to disk and ultimately enhance analysis efficiency, we propose the addition of
a level-three software trigger to the \gx~data acquisition system.  We should note that the 
\gx~detectors and data acquisition architecture are designed to handle a rate of $10^8~\gamma/$s.
However, the optimum photon flux for taking data will depend on the event 
reconstruction at high luminosity and needs to be studied under realistic experimental conditions.
If our extraction of amplitudes is not statistics limited, we may optimize
the flux to reduce systematic errors.

\subsection{Kaon identification in \gx}

\begin{figure*}
\begin{center}
\includegraphics[width=\linewidth]{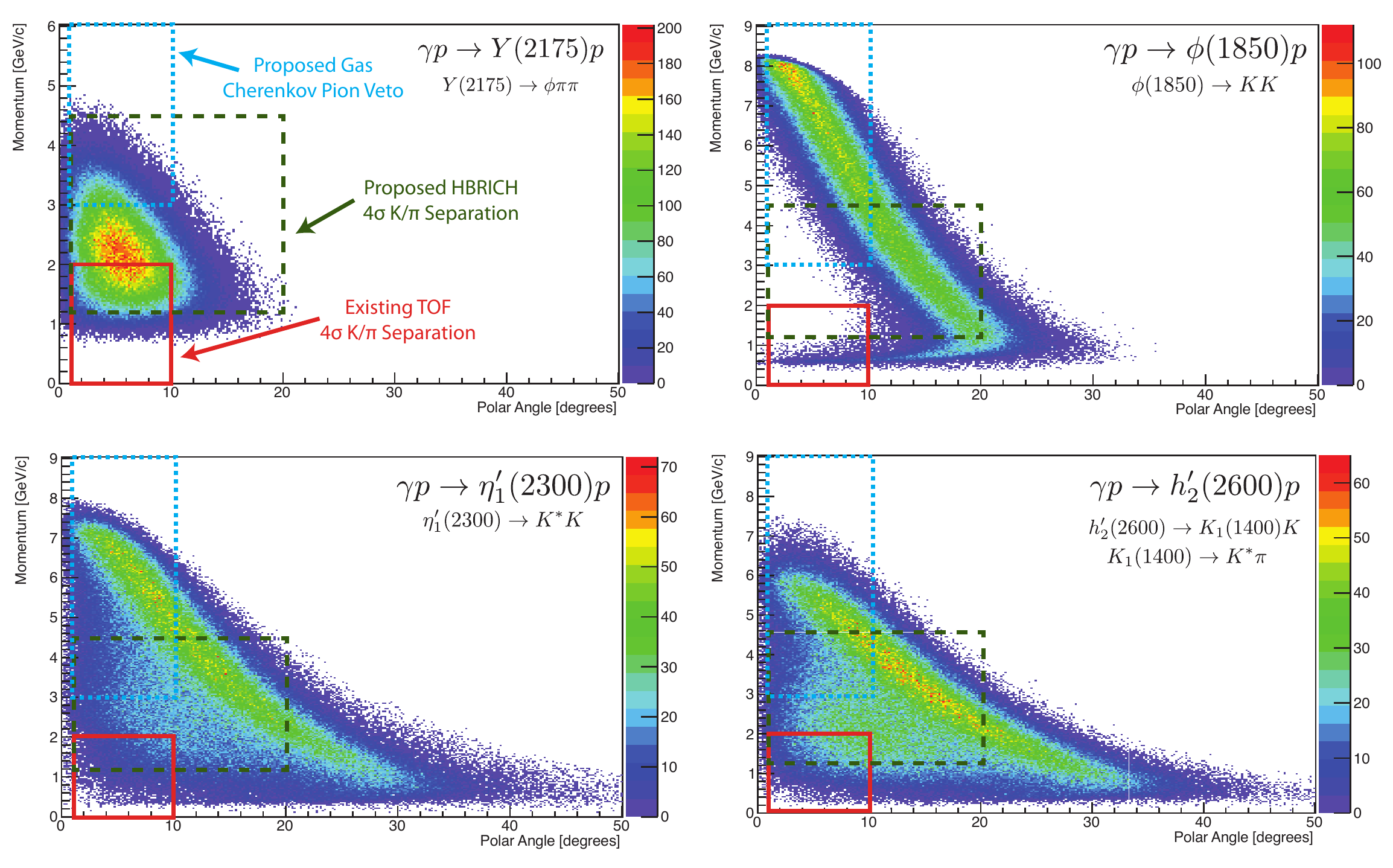}
\caption{\label{fig:kaon_kin}Plots of particle density as a function of momentum and polar angle 
for all kaons in a variety of different production channels.  Shown in solid (red), dashed (green), and dotted (blue) are 
the regions of phase space 
where the existing time-of-flight (TOF) detector, proposed HBRICH, and proposed gas Cherenkov provide
pion/kaon discrimination at the four standard deviation level.}
\end{center}
\end{figure*}

In order to understand how decays of heavy exotic $s\bar{s}$ mesons might populate
the \gx~detector, we simulated several channels noted in Table~\ref{tab:exotic_modes}.
Each was simulated with a production cross section proportional to $e^{-5|t|}$, where $|t|$ 
is the four-momentum transfer squared from the photon to the
nucleon\footnote{For $t$ production coefficients less (greater) than $5$~GeV$^{-2}$, we expect the acceptance to
increase (decrease).  The true $t$ dependence is unknown for the reaction of interest; therefore, 
we use 5~GeV$^{-2}$ as a conservative estimate.}.  Figure~\ref{fig:kaon_kin} 
shows how four key reactions populate the space of
kaon momentum and polar angle.  Assuming a momentum resolution of 2\%, a path length resolution
of 3~cm, and a timing resolution of 70~ps, the existing forward TOF detector provides four standard deviations
of separation between kaons and pions in the region of $1^\circ-10^\circ$ and $<2.0~$GeV/$c$
(indicated by the red box in Figs.~\ref{fig:kaon_kin}).  One can see that the system
is clearly inadequate for efficiently detecting kaons produced in these key reactions.

\begin{figure}
\begin{center}
\includegraphics[width=\linewidth]{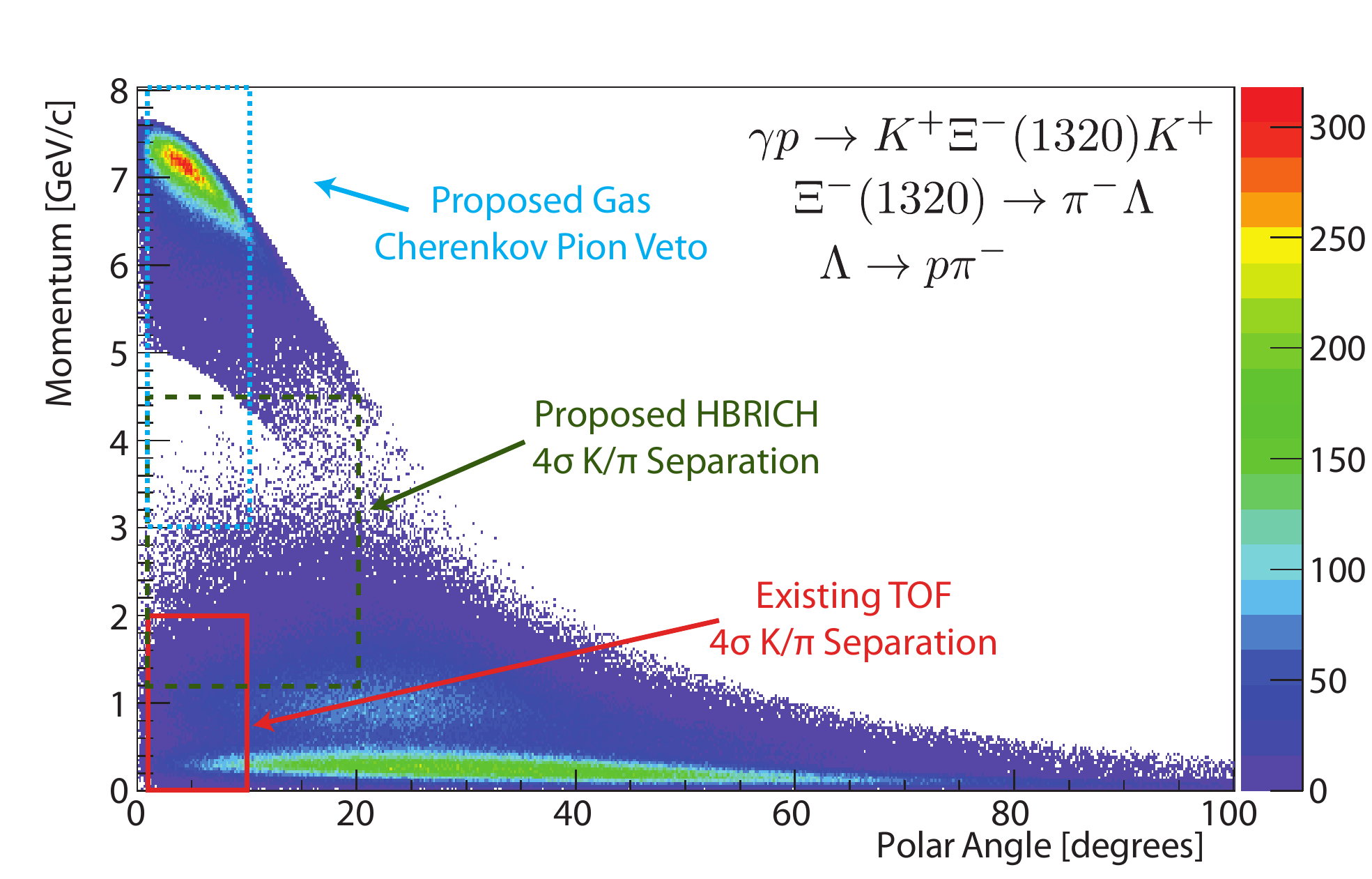}
\includegraphics[width=\linewidth]{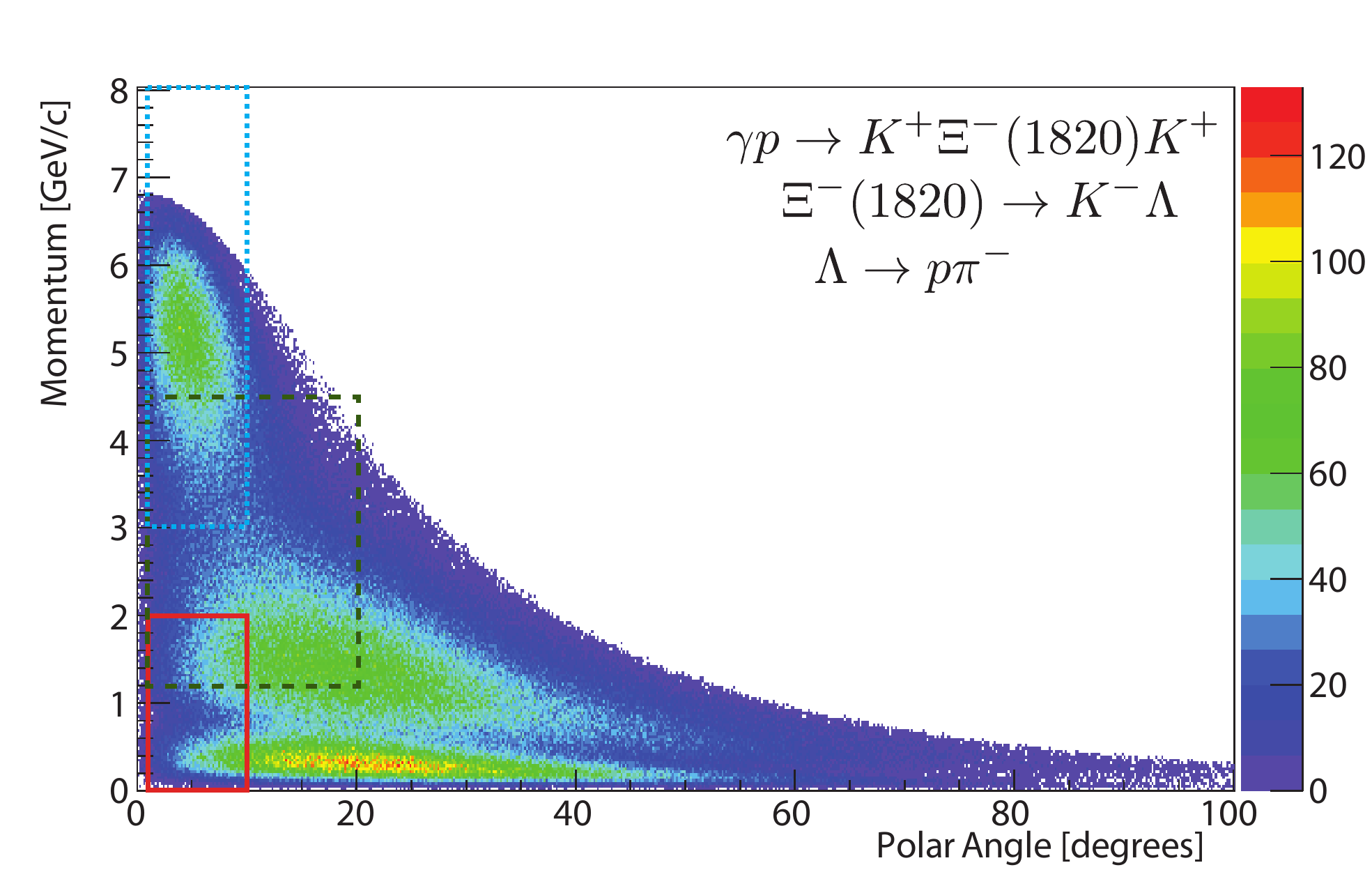}
\caption{\label{Figure:Production} Generated momentum versus polar
    angle for all tracks in the simulated reactions (top) $\gamma p\to
    K^+\,\Xi^-(1320) K^+$ and (bottom) $\gamma p\to K^+\,\Xi^-(1820)
    K^+$. The three high-density regions in each plot are populated with,
    from lowest to highest momentum, pions, kaons and protons, and kaons.
    Regions of coverage for various particle identification detectors are indicated.} 
\end{center}
\end{figure}

We have also simulated the production of the $\Xi^-(1320)$ and
$\Xi^-(1820)$~resonances to better understand the kinematics of these
reactions. The photoproduction of the $\Xi^-(1320)$ decaying to
$\pi^-\Lambda$ and of the $\Xi^-(1820)$ decaying to $\Lambda K^-$ are
shown in Fig.~\ref{Figure:Production}. These reactions result in
$K^+K^+\pi^-\pi^-p$ and $K^+K^+K^-\pi^-p$ final states,
respectively. Reactions involving excited Cascades have ``softer''
forward-going kaons, and there is more energy available on average to
the Cascade's decay products. Both plots show three regions of high
density. The upper momentum regions ($>4$~GeV/$c$) consist of
forward-going $K^+$ tracks from the associated production of an
excited hyperon. The middle momentum regions (0.6-2.0~GeV/$c$) are a
mixture of kaon and proton tracks, while the lower regions
($<0.6$~GeV/$c$) contain mostly $\pi^-$~tracks. The
kaon tracks with momenta larger than about 2.0~GeV/$c$ cannot be positively
identified with the current \gx~PID system. Accurate particle
identification will be essential for rejection of background
events. The addition of a forward kaon identification system will allow for
the separation of fast, forward-going kaons from pions, cleaning up
the event sample.

Two types of Cherenkov detectors that are commonly used for particle identification are the threshold
Cherenkov detector and the ring-imaging Cherenkov (RICH) detector.  In a threshold Cherenkov detector,
a radiator is chosen such that, at a fixed momentum, pions emit Cherenkov radiation, but heavier
kaons do not.  The kaon identification is therefore performed by assuming that charged tracks that
are not pions are kaons.  The goal of a RICH is to image the Cherenkov cone emitted by all charged particles
traversing the radiator.  By using a thin radiator, the emitted Cherenkov light projects an elliptical ring onto the
detector.  The measured Cherenkov angle along with the momentum can be used to infer
the particle mass.  The RICH is capable of positively identifying all three common species of hadrons:  pions, 
kaons, and protons.  

Below we discuss conceptual designs for both types of detectors.  Neither of the two designs presented
are final -- the \gx~Collaboration is actively working on design and simulation of detectors in order to develop
a forward particle identification solution that is optimized for the \gx~physics program.  

\subsubsection{Threshold gas Cherenkov}

The initial \gx~proposal included the construction of a threshold gas Cherenkov detector in between the 
solenoid and the forward time-of-flight systems.  A single C$_4$F$_{10}$ gas
radiator with an index of refraction $n= 1.0015$ was proposed, which has a pion momentum threshold
for Cherenkov radiation at 2.5~GeV/$c$.  The design utilized two sets of focussing mirrors to
reflect the light into phototubes.  A conceptual drawing along with proposed placement in the
\gx~detector is shown in Fig.~\ref{fig:pid}.  If the threshold for detection is set at about five photoelectrons, then
the proposed detector would be capable of providing a pion veto above 3~GeV/$c$ momentum
in the region less than $10^\circ$.  This detector would greatly improve the kaon detection efficiency over
the time-of-flight system; however, detection would still be limited to the forward direction, and a gap in
detection capability exists for kaon momenta between 2.0 and 3.0~GeV/$c$.  In addition, no separation
between kaons and protons is provided, as neither will generate Cherenkov radiation in C$_4$F$_{10}$ in the momentum
range of interest.  (Kaons begin to radiate around 9~GeV/$c$.) 
A secondary radiator, {\it e.g.} aerogel, may enhance the performance.  
A final concern is the background rate in the detector.  We know that
the dominant background is $e^+e^-$ pairs generated by interactions of the photon beam with material.
Detectors near the \gx~beamline, {\it e.g.,} the FCAL, FDC, and TOF, are expected to experience
electromagnetic background at the MHz level.  Significant backgrounds may lead to false rejection of
true kaons.  For comparison, the region of kaon phase space where the threshold gas Cherenkov detector
would effectively operate as a pion veto is indicated in Figs.~\ref{fig:kaon_kin} and~\ref{Figure:Production} with a 
blue dotted line.

\subsubsection{Hadron blind RICH}

\begin{figure*}
\begin{center}
\includegraphics[width=\linewidth]{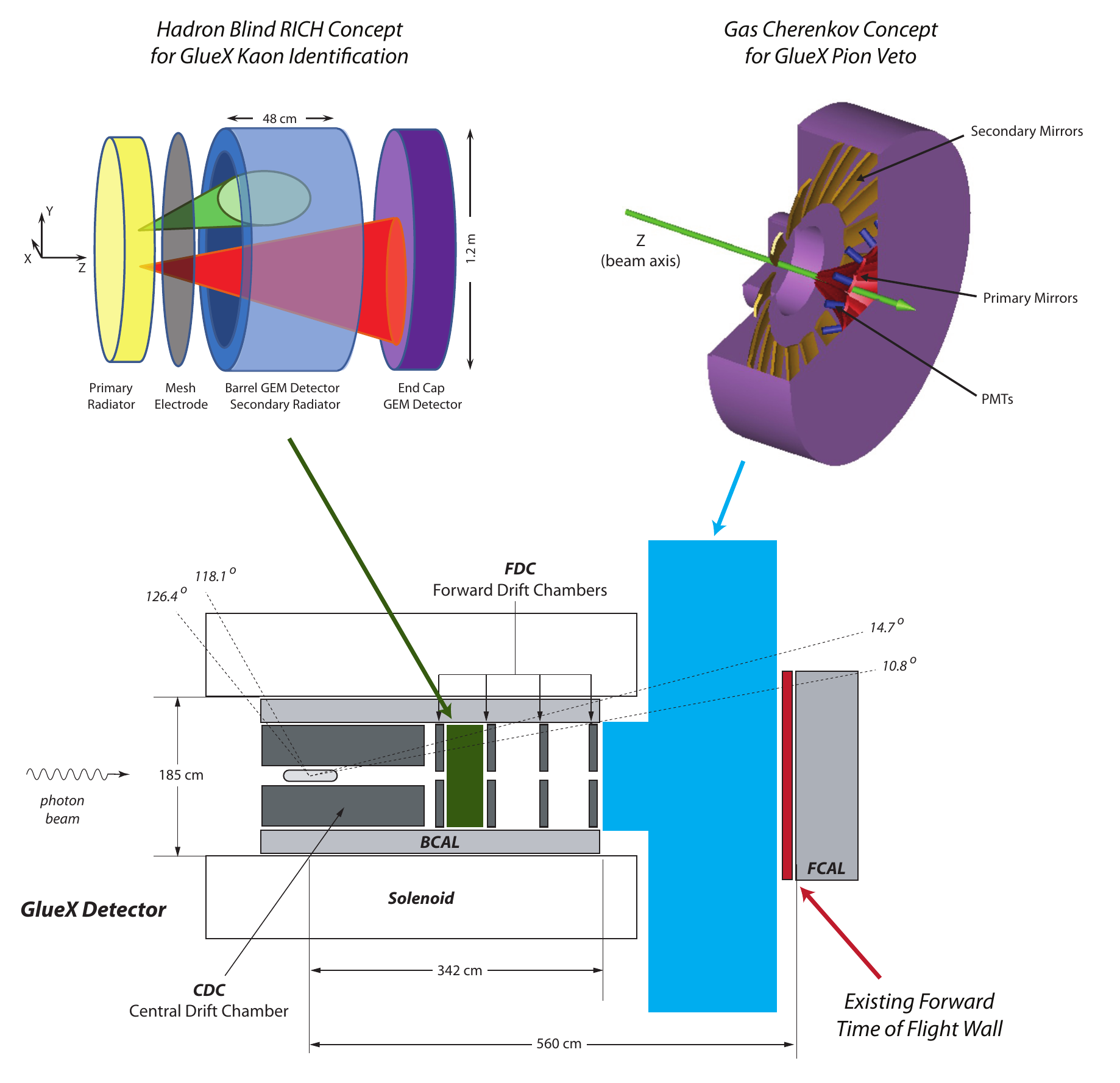}
\caption{\label{fig:pid}Conceptual drawings of both the hadron blind RICH (HBRICH) and threshold gas Cherenkov detectors
and their proposed locations within the \gx~detector.}
\end{center}
\end{figure*}

With the goal of having positive kaon identification up to momenta of 4.5~GeV/$c$ and enhancing the
angular coverage up to $20^\circ$, a conceptual design for a ``hadron blind RICH" (HBRICH) detector
for \gx~has been developed.  The detector is inspired by the existing proximity
focusing RICH detector in use in Hall~A, which utilizes CsI to convert the Cherenkov
photons to electrons~\cite{Iodice:2005ex}.  In order to make the detector blind to ionization
produced by hadrons traversing the expansion volume, a reverse bias is applied to clear the volume 
of ionization electrons.  Such a technique was pioneered by the PHENIX hadron blind detector~\cite{Anderson:2011jw}.
A sketch of the detector, along with its proposed placement, is shown in Fig.~\ref{fig:pid}.
The upstream end cap contains the primary radiator, followed by a mesh electrode to provide reverse bias. 
The barrel of the detector (1.9 m$^2$) and the downstream end cap (1.1 m$^2$) will be covered 
by three-stage gas electron multiplier (GEM) foils~\cite{Sauli:1997qp} together with pad readout that records the
image charge of the electron avalanche on the GEM. The center 
volume can be filled with gas as the secondary radiator to provide complementary PID capability and enhance the
range of momenta covered by the detector.

We can make a rough estimate of the number of readout channels needed for the design
described above.  An $n=1.1$ radiator, {\it e.g.}, aerogel, provides angular separation of kaons and pions 
within the momentum range of interest.  At 4.5 GeV$/c$, the Cherenkov angles of kaons and 
pions differ by about 14~mrad after photons are refracted 
from the radiator into the detector volume. Assuming an average of 10 photoelectrons per track, 
in order to reach a 4$\sigma$ separation, the maximum size for the readout pads is about 19~mm.
A 15~mm$~\times~$15~mm pad is sufficient for the end cap.  The barrel readout may need
double the resolution in one direction to accommodate the elliptical projection of the rings.
The total number of channels would be approximately 22,000,
which is of a comparable scale to the number of channels in the \gx~FDC.  The projected momentum
and angular coverage of the HBRICH is shown in Figs.~\ref{fig:kaon_kin} and~\ref{Figure:Production} with a 
green dashed line.

In comparison to the gas threshold Cherenkov, the design and development of an HBRICH
is notably more challenging.  While portions of the design are inspired by existing detectors,
a detector like the HBRICH has never been constructed.  The detector would require the insertion of a large
amount of material into the active tracking volume, which could have adverse effects on tracking
resolution and calorimeter efficiency.  While there is a lot of activity in the development of large-area GEM
foils~\cite{Alfonsi:2010zz} driven by the desire to build high-resolution time projection chambers, 
GEM-based detectors of the scale of the proposed HBRICH have not yet been constructed.  
In addition, significant work is needed to develop a large scale GEM readout for such a detector.  
The gas threshold Cherenkov, on the other hand, is
based primarily of proven technology that has been used in many detectors for decades.
Either detector would provide a significant enhancement in the kaon identification capability of \gx~and 
greatly expand the discovery potential of the experiment.  

The collaboration is also exploring other options.  For example, a dual radiator (aerogel and C$_4$F$_{10}$) 
RICH, similar to that developed for LHCb~\cite{Pickford:2009zz}, could fit in the region of the proposed
threshold gas Cherenkov detector.  Assuming the same performance demonstrated by LHCb is achievable
in \gx, this detector would provide positive kaon and pion identification for particles with momenta greater
than 1~GeV/$c$; however, its angular coverage would be limited to the region $<10^\circ$ about the beam axis.
As noted above, the collaboration has an active program of detector design ongoing and is expected to 
reach a final design decision within the next six months.

\subsection{Desired beam time and intensity}
\label{sec:beamtime}

One can estimate the total number of observed events in some stable 
final state $N_i$ using the following equation.
\begin{equation}
N_i = \epsilon_i \sigma_i n_\gamma n_t T,
\label{eq:rate}
\end{equation}
where $\epsilon_i$ and $\sigma_i$ are the detection efficiency and 
photoproduction cross section of the final state $i$, $n_\gamma$ is the rate 
of incident photons on target, $n_t$ is the number of scattering centers per 
unit area, and $T$ is the integrated live time of the detector.  For a 30~cm 
LH$_2$ target, $n_t$ is 1.26~b$^{-1}$.  (A useful rule of thumb is that at 
$n_\gamma = 10^7~\gamma$/s a 1~$\mu$b cross section will result in the 
production of about $10^{6}$ events per day.)  It is difficult to estimate the 
production cross section for many final states since data in the \gx~energy 
regime are sparse.  (For a compendium of photoproduction cross sections, 
see Ref.~\cite{Baldini:1988ti}.)  Table~\ref{tab:yields} lists key final states
for initial exotic hybrid searches along with assumed production cross sections\footnote{Some estimates are 
based on actual data from Ref.~\cite{Baldini:1988ti} for cross sections at a similar beam energy, 
while others are crudely estimated from the product of branching ratios of heavy meson decays, {\it i.e.,} a proxy 
for light meson hadronization ratios, and known photoproduction cross sections.}.

In order to estimate the total yield in Phase~II and~III running, we assume 90 PAC days
of beam\footnote{We plan to utilize 30 of the 120 approved PAC days for the Phase~I commissioning
of the detector.} and estimate that 80\% of the delivered beam will be usable for analysis 
(about $6\times10^6$~s total) at an intensity of $n_\gamma = 10^7~\gamma$/s in the coherent bremsstrahlung peak.
For the proposed Phase~IV running, we assume 200 PAC days of beam with 80\% usable for physics analysis
and an {\em average intensity} of $5\times10^7~\gamma/s$.  We assume the detection 
efficiency for protons, pions, kaons, and photons to be 70\%, 80\%, 70\%, and 80\%, respectively.  
Of course, the true efficiencies are dependent on 
software algorithms, kinematics, multiplicity, and other variables; however, the dominant uncertainty 
in yield estimates is not efficiency 
but cross section.  These assumed efficiencies reproduce signal selection efficiencies in detailed simulations of
 $\gamma p \to \pi^+\pi^-\pi^+ n$, $\gamma p \to \eta \pi^0 p$, $\gamma p \to b_1^\pm \pi^\mp p$, 
and $\gamma p \to f_1\pi^0 p$ performed by the collaboration.

\begin{table}
\begin{center}
\caption{\label{tab:yields}A table of hybrid search channels, estimated cross sections, and approximate numbers of observed events.  See text for a discussion of the underlying assumptions.  The subscripts on $\omega$, $\eta$, and $\eta^\prime$ indicate the decay modes used in the efficiency calculations.  If explicit charges are not indicated, the yields represent an average over various charge combinations.}
\begin{tabular}{cccc}\hline\hline 
& Cross & Approved & {\em Proposed} \\
Final &  ~~Section~~ & Phase II and III & Phase IV \\
  State & ($\mu$b) & ($\times10^6$ events) & ($\times10^6$ events) \\  \hline
$\pi^+\pi^-\pi^+$ & 10 & 300 & 3000 \\
$\pi^+\pi^-\pi^0$ & 2 & 50 & 600 \\
$KK\pi\pi$ & 0.5 & -- & 100 \\
$\omega_{3\pi}\pi\pi$ & 0.2 & 4 & 40 \\
$\omega_{\gamma\pi}\pi\pi$ & 0.2 & 0.6 & 6 \\
$\eta_{\gamma\gamma}\pi\pi$ & 0.2 & 3 & 30 \\
$\eta_{\gamma\gamma}\pi\pi\pi$ & 0.2 & 2 & 20 \\
$\eta^\prime_{\gamma\gamma}\pi$ & 0.1 & 0.1 & 1 \\
$\eta^\prime_{\eta\pi\pi}\pi$ & 0.1 & 0.3 & 3 \\
$KK\pi$ & 0.1 & -- & 30 \\ \hline \hline
\end{tabular}
\end{center}
\end{table}

\begin{figure}
 \begin{center}
 \includegraphics[width=\linewidth]{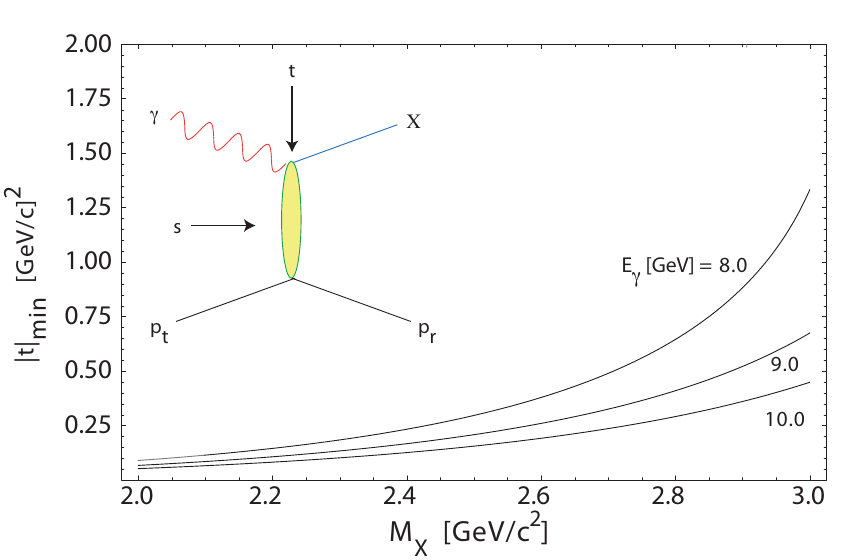}
 \caption{\label{fig:tmin}Dependence of $|t|_\mathrm{min}$ on the mass of the outgoing meson system $M_X$.  The lines indicate incident photon energies of 8.0, 9.0, and 10.0 GeV.}
\end{center}
 \end{figure}

\begin{figure}
 \begin{center}
 \includegraphics[width=\linewidth]{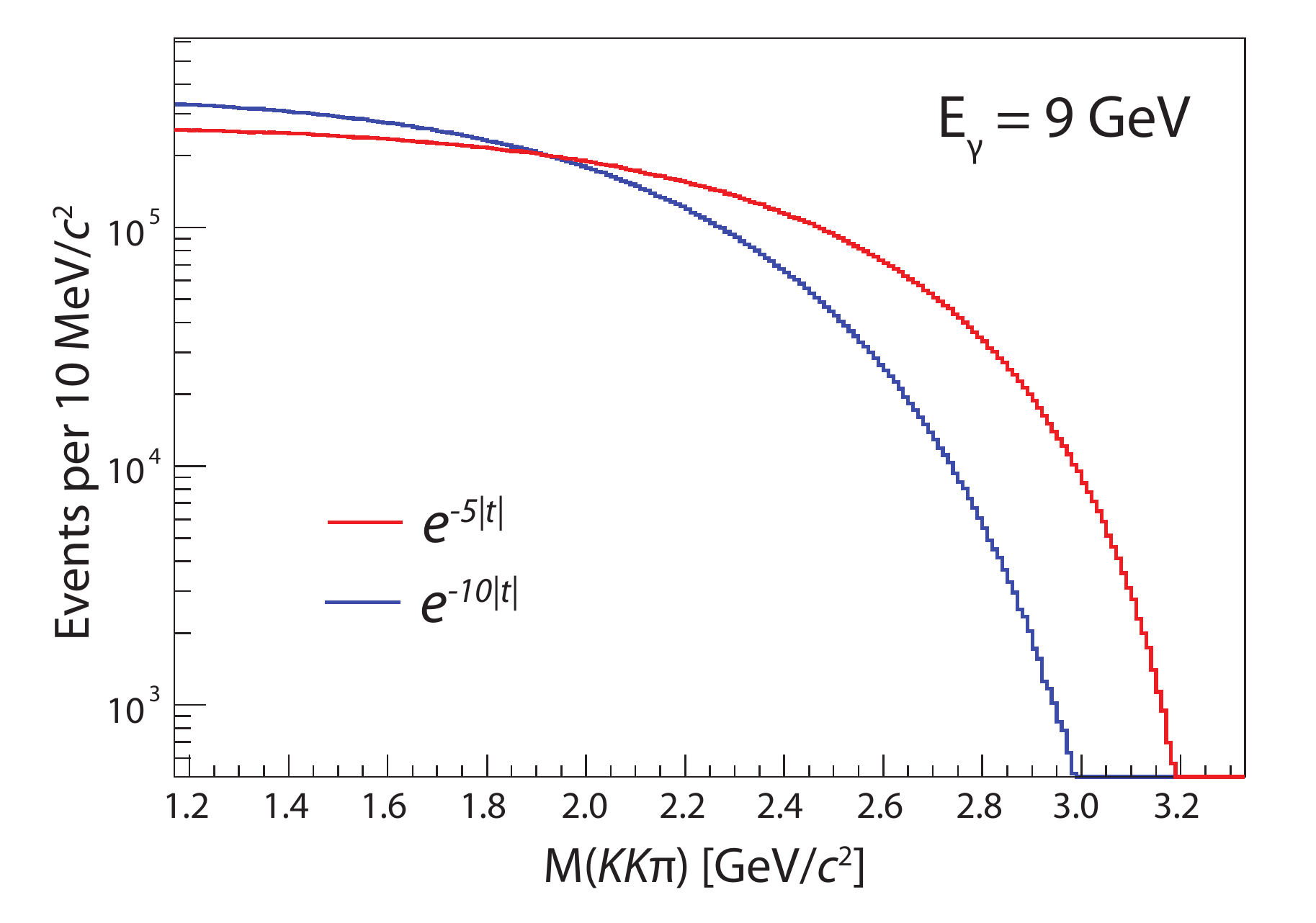}
 \caption{\label{fig:n_vs_mass}A figure showing the number of expected events per 10~MeV/$c^2$ bin in $KK\pi$ invariant mass, integrating over all allowed values of $t$, and assuming $3\times10^7$ events in total are produced.  No dependence on $M(KK\pi)$ is assumed, although, in reality, the mass dependence will likely be driven by resonances.  Two different assumptions for the $t$ dependence are shown.  The region above 2.5~GeV/$c^2$ represents about 8\% (2\%) of all events for the $\alpha = 5 (10)$~GeV$^{-2}$ values.}
\end{center}
 \end{figure}

Photoproduction of mesons at 9~GeV proceeds via peripheral production (sketched in the inset of Fig.~\ref{fig:tmin}).
The production can typically be characterized as a function of $t\equiv (p_X-p_\gamma)^2$, with the production 
cross section proportional to $e^{-\alpha|t|}$.  The value of $\alpha$ for 
measured reactions ranges from 3 to 10~GeV$^{-2}$.  
This $t$-dependence, which is unknown for many key \gx~reactions, 
results in a suppression of the rate at large values of $|t|$,
which, in turn, suppresses the production of high mass mesons.  Figure~\ref{fig:tmin} shows the minimum value of 
$|t|$ as a function of the produced meson mass $M_X$ for a variety of different photon energies.  The impact
of this kinematic suppression on a search for heavy states is illustrated in Figure~\ref{fig:n_vs_mass}, 
where events are generated according to the $t$ distributions with both $\alpha=5~$GeV$^{-2}$ and
 10~GeV$^{-2}$ and uniform in $M_X$.  Those that are kinematically 
 allowed ($|t|>|t|_\mathrm{min}$) are retained.  The $y$-axis indicates the number of events 
 in 10~MeV/$c^2$ mass bins, integrated over the allowed region in $t$, and assuming a 
 total of $3\times 10^7$ events are collected.  The region above $M_X=2.5~$GeV$/c^2$, where one would want to search
 for states such as the $h_2$ and $h_2^\prime$, contains only about 5\% of all events due to the suppression
 of large $|t|$ that is characteristic of peripheral photoproduction.
  
There are several considerations in determining how much data one needs in any particular final state.
In order to perform an amplitude analysis of the final state particles, one 
typically separates the data into bins of momentum transfer $t$ and resonance mass $M_X$.  
The number of bins in $t$ could range from one to greater than ten, depending on the
statistical precision of the data; a study of the $t$-dependence, if statistically permitted, provides valuable information on the
production dynamics of particular resonances.  One would like to make the mass bins as small as possible
in order to maximize sensitivity to states that have a small total decay width; however, it is not
practical to use a bin size that is smaller than the resolution on $M_X$, which is of order 10~MeV/$c^2$.
In each bin of $t$ and $M_X$, one then needs enough events to perform an amplitude analysis, which is about $10^4$.
Figure~\ref{fig:n_vs_mass} demonstrates that, under some assumptions about the production, this
level of statistics is achievable for the mass region where $s\bar{s}$ exotics are expected to reside.  
More importantly, it stresses the need to maximize the statistical precision of the data in order to 
be able to search for heavier exotic mesons.

Recent CLAS data for the $\Xi(1320)$ are consistent with $t$-slope values ranging from 
1.11 to 2.64~$({\rm GeV}/c)^{-2}$ for photon energies between 2.75 and 3.85~GeV~\cite{Guo:2007dw}. 
Values for excited Cascades are not well known, but a recent unpublished CLAS analysis 
of a high-statistics data sample (JLab proposal E05-017) indicates that the $t$-slope value 
flattens out above 4~GeV at a value of about 1.7~$({\rm GeV}/c)^{-2}$~\cite{GoetzThesis}. 
We have used a value of 1.4~$({\rm GeV}/c)^{-2}$ for the $\Xi^-(1320)$ and 1.7~$({\rm GeV}/c)^{-2}$ 
for the $\Xi^-(1820)$ in our simulations at 9 GeV. For the reactions in Fig.~\ref{Figure:Production}, we 
estimate a detection efficiency of about 10\%. The most recent CLAS analysis 
has determined a total cross section of about 8~nb and 2~nb for the $\Xi^-(1320)$ 
and $\Xi^-(1530)$, respectively, at $E_\gamma=5$~GeV. The analysis shows that the 
total cross section levels out above 3.5~GeV, but the energy range is limited at 
5.4~GeV~\cite{GoetzThesis}. A total number of about 20,000 $\Xi^-(1320)$~events 
was observed for the energy range $E_\gamma\in [2.69,\,5.44]$~GeV. \gx~will 
collect over an order of magnitude more data:  using Eq.~(\ref{eq:rate}), 
we expect about 400,000 $\Xi^-(1320)$ and 100,000 $\Xi^-(1530)$ events to be collected in Phase IV.

In summary, we request a production run consisting of 200 days beam time at an average 
intensity of $5\times 10^7~\gamma$/s for production 
Phase~IV running of the \gx~experiment.  An additional 20 days of beam will be needed to commission the
Phase~IV detector hardware, with some time between the 20 day commissioning run and the 200-day Phase IV run
to assess the results.  It is anticipated that the Phase~IV
intensity will start around $10^7~\gamma/$s, our Phase~III intensity, and increase toward the \gx~design goal 
of $10^8~\gamma/$s as we understand the detector capability for these high rates based on the 
data acquired at $10^7~\gamma/$s.  The data sample acquired will provide an order of magnitude improvement
in statistical precision over the initial Phase II and III running of \gx, which will allow a detailed study of high-mass states.

\begin{table*}
\begin{center}
\caption{\label{tab:params}A table of relevant parameters for the various phases of \gx~running.}
\begin{tabular}{l|ccc|c}\hline\hline
 & & Approved & & {\em Proposed} \\
 & Phase I & Phase II & Phase III & ~~Phase IV~~ \\ \hline 
 Duration (PAC days) & 30 & 30 & 60 & 220\footnote{Twenty days are dedicated to Phase~IV hardware commissioning and will be scheduled in advance of the 200-day physics run.} \\
 Minimum electron energy (GeV) & 10 & 11 & 12 & 12 \\
 Average photon flux  ($\gamma$/s) & $10^6$ & $10^7$ & $10^7$ & $5 \times 10^7$ \\
 Average beam current (nA) & 50 - 200\footnote{An amorphous radiator may be used for some commissioning and later
 replaced with a diamond.}  & 220 & 220 & 1100 \\
 Maximum beam emittance (mm$\cdot\mu$r) & 50 & 20 & 10 & 10 \\
 Level-one (hardware) trigger rate (kHz) & 2 & 20 & 20 & 200 \\
 Raw Data Volume (TB) & 60 & 600 & 1200 & 2300\footnote{This volume assumes the implementation of the 
 proposed level-three software trigger.} \\ \hline\hline
 \end{tabular}
 \end{center}
 \end{table*}

\subsection{Level-three trigger}

The energy spectrum of photons striking the target ranges from near zero to the full 12~GeV incident electron
energy.  For physics analyses, one is primarily interested in only those events in the coherent peak
around 9~GeV, where there is a signal in the tagger that determines the photon energy.  At a rate of $10^7~\gamma$/s,
the 120~$\mu$b total hadronic cross section at 9~GeV corresponds to a tagged hadronic event rate of
about 1.5~kHz.  Based on knowledge of the inclusive photoproduction cross section as a function of energy,
calculations of the photon intensity in the region outside the tagger acceptance, and estimates for the 
trigger efficiency, a total trigger rate of about 20~kHz is expected.  At a typical raw event size of 15~kB, the
expected data rate of 300~MB/s will saturate the available bandwidth to disk -- rates higher than $10^7~\gamma$/s
cannot be accommodated with the current data acquisition configuration.

For the high-intensity running, we propose the development of a level-three software trigger
to loosely skim events that are consistent with a high energy $\gamma p$ collision.  The
events of interest will be characterized by high-momentum tracks and large energy deposition in
the calorimeter.  Matching observed energy with a tagger hit is a task best suited for software algorithms
like those used in physics analysis.  It is expected that a processor farm can analyze multiple events in parallel,
providing a real time background rejection rate of at least a factor of ten.  While the exact network topology
and choice of hardware will ultimately depend on the speed of the algorithm, the system will need to
accommodate the 3~GB/s input rate, separate data blocks into complete events, and output the accepted
events to disk at a rate of $<300$~MB/s.  The software trigger has the added advantage of increasing
the concentration of tagged $\gamma p$ collision events in the output stream, which better optimizes use of disk
resources and enhances analysis efficiency.

A simple estimate indicates that the implementation of a level-three trigger easily results in a net cost
savings rather than a burden.  Assuming no bandwidth limitations, if we write the entire unfiltered 
high-luminosity data stream to tape, the
anticipated size is about 30 petabytes per year\footnote{This is at the \gx~design intensity of $10^8~\gamma/$s, which is
higher than our Phase IV average rate of $5\times10^7$ by a factor of two; however, other factors that would
increase the data volume, such as event size increases due to higher-than-estimated noise or the additional
data size associated with a RICH detector, have not been included.}.  
Estimated media costs for storage of this data at the time of running would
be \$300K, assuming that no backup is made.  A data volume of this size would require the 
acquisition of one or more additional tape silos at a cost of about \$250K each.  Minimum storage costs
for a multi-year run will be nearing one million dollars.  Conversely if we assume a level-three trigger algorithm
can run a factor of ten faster than our current full offline reconstruction, then we can process events at a 
rate of 100~Hz per core.  The anticipated peak high luminosity event rate of 200~kHz would require 2000 cores, which
at {\em today's costs} of 64-core machines would be about \$160K.  Even if a factor of two in computing is added to
account for margin and data input/output overhead, the cost is significantly lower than the storage cost.  
Furthermore, it is a fixed cost
that does not grow with additional integrated luminosity, and it reduces the processing cost of the final stored data set 
when a physics analysis is performed on the data.

\subsection{Phase IV hardware development plan}
 
The collaboration has started a program to develop a detailed and final conceptual design for a forward
particle identification system.  As demonstrated, the existing \gx~software framework will
permit a simulation of reactions of interest in addition to providing the capability to analyze the impact
of additional detector material on existing calorimeter and tracking performance.  The development
of a kaon detector will be a multi-institutional effort.  Currently, simulations of kaon production are being
carried out at Florida State University.  Jefferson Lab has contributed to optimization of the mirror design
for the gas Cherenkov detector in addition to the conceptual development of the HBRICH.  This year,
Massachusetts Institute of Technology, a group that has experience with the RICH detector at LHCb,
has joined the collaboration with the desire to take a lead role in the construction of a forward particle identification system.
A workshop has been organized for May 2012 in order to develop a plan for moving towards a complete design.  The 
collaboration has set a goal of converging on a design decision by the fall of 2012 so that an accurate
cost estimate can be obtained.  It is expected that key proponents within the collaboration will initially
seek funding through appropriate NSF and DOE programs, {\it e.g.,} the NSF Major Research Instrumentation 
Program or the DOE Early Career Award Program, to fund the construction of the detector.  It should be noted
that the member institutions of \gx~have significant detector construction capabilities -- most major detector systems
have been constructed offsite at \gx~institutions.  As construction of the baseline \gx~detector
will be complete at the institutions by 2014, resources may be redirected to assist in the construction of a forward
particle identification system.

The level-three trigger can largely be developed as proposed in the initial \gx~design.  The present baseline
data acquisition system has been carefully developed so that a level-three software trigger can be
easily accommodated in the future.  Given the standardized approach to data acquisition and networking
in place at Jefferson Lab, the lab staff would have to play a key role in developing and deploying the 
level-three trigger.  As noted above, at the \gx~design intensity of $10^8~\gamma/$s, the operational cost
of managing all of the data produced by the experiment will likely outstrip the one-time hardware cost
associated with developing a level-three trigger.

\section{Summary}

In summary, we propose an expansion of the present baseline \gx~experimental program to include
both higher statistics and the capability to detect kaons over a broad range of phase space.
This detection capability would allow the identification of meson states with an $s\bar{s}$ component and
the search for doubly-strange $\Xi$-baryon states.  The program would require 220 days of beam time 
with  9~GeV tagged photons at an average intensity
of $5\times 10^7~\gamma/s$.  The data taking will be divided into an initial twenty-day commissioning period 
for the Phase~IV hardware and then a 200-day production Phase~IV physics run.  
The Phase~IV running will provide an increase in statistics over the approved Phase~II and III
\gx~runs by an order of magnitude.  To execute this program, a new forward particle identification
system will be developed in order to supplement the existing capability of the
forward time-of-flight system.  In addition, a level-three software trigger is needed to provide
an order of magnitude reduction in the rate at which background photoproduction events are written
to disk, which would allow the experiment to run efficiently at high intensity and minimize the final 
disk and tape resources needed to support the data.  With these enhanced capabilities, \gx~will
be able to conduct an exhaustive study of both $s\bar{s}$ and $\ell\bar{\ell}$ mesons, thereby providing
crucial experimental data to test the quantitative predictions of hybrid and conventional mesons that
are emerging from both lattice calculations and other models of QCD.

\end{document}